\pdfoutput=1

\documentclass[hyperindex,breaklinks=true, colorlinks, citecolor=blue]{aa}  
\usepackage{graphicx}
\usepackage{hyperref}

\usepackage{txfonts}
\usepackage{aas_macros}
\usepackage{subeqnarray}
\usepackage{natbib}
\usepackage{version}
\usepackage{ulem}

\def\ie{\emph{i.e.\/}}

\def\eg{\emph{e.g.\/}}

\newlength{\thsize}
\newlength{\hhsize}
\newlength{\qhsize}
\begin{document}

\title{Interplay of dust alignment, grain growth and magnetic fields in polarization: lessons from the emission-to-extinction ratio}


   \author{L. Fanciullo, \inst{1}
        V. Guillet, \inst{1}	
	F. Boulanger, \inst{1}
          \and
	A. P. Jones \inst{1}
          }


   \institute{Institut d'Astrophysique Spatiale (IAS), B\^atiment 121, Universit\'e Paris-Sud 11 and CNRS, F- 91405 Orsay, France}

   \date{}

 
  \abstract
{Polarized extinction and emission from dust in the interstellar medium (ISM) are hard to interpret, as they have a complex dependence on dust optical properties, grain alignment and magnetic field orientation. This is particularly true in molecular clouds. The aforementioned phenomena are usually considered independently in polarization studies, while it is likely that they all contribute and their effects have yet to be disentangled. }
{The data available today are not yet used to their full potential. The combination of emission and extinction, in particular, provides information not available from either of them alone. We combine data from the scientific literature on polarized dust extinction with \Planck\ data on polarized emission and we use them to constrain the possible variations in dust and environmental conditions inside molecular clouds, and especially translucent lines of sight, taking into account magnetic field orientation. }
 {We focus on the dependence between $\lmax$ -- the wavelength of maximum polarization in extinction -- and other observables such as the extinction polarization, the emission polarization and the ratio of the two. We set out to reproduce these correlations using Monte-Carlo simulations where the relevant quantities in a dust model -- grain alignment, size distribution and magnetic field orientation -- vary to mimic the diverse conditions expected inside molecular clouds. }
 {None of the quantities chosen can explain the observational data on its own: the best results are obtained when all quantities vary significantly across and within clouds. 
 However, some of the data -- most notably the stars with low emission-to-extinction polarization ratio -- are not reproduced by our simulation. }
{Our results suggest not only that dust evolution is necessary to explain polarization in molecular clouds, but that a simple change in size distribution is not sufficient to explain the data, and point the way for future and more sophisticated models.}

\def\eg{\textit{e.g.}}
\def\ie{\textit{i.e.}}
\def\Planck{\textit{Planck}}
\def\IRAS{\textit{IRAS}}
\def\nside{N_{\rm side}}

\def\Psiex{\psi_{\rm ext}}
\def\Psiem{\psi_{\rm 353}}
\def\sPsiex{\sigma(\psi_{\rm ext})}
\def\sPsiem{\sigma(\psi_{\rm 353})}
\def\Pem{P_{\rm 353}}
\def\sPem{\sigma(P_{\rm 353})}
\def\Iem{I_{\rm 353}}
\def\tem{\tau_{\rm 353}}
\def\tv{\tau_{\rm V}}
\def\pv{p_{\rm V}}
\def\lmax{\lambda_{\rm max}}
\def\pmax{p_{\rm max}}
\def\Rpp{\Pem/\pv}
\def\Rv{R_{\rm V}}
\def\Av{A_{\rm V}}
\def\EBV{E(B-V)}
\def\Nh{N_{\rm H}}

\def\fmax{f_{\rm max}}
\def\stiff{s}
\def\athresh{a_{\rm alig}}
\def\amax{a_{\rm max}}
\def\Cext{C_{\rm ext}}
\def\Cabs{C_{\rm abs}}
\def\Csca{C_{\rm sca}}
\def\Cpol{C_{\rm pol}}
\def\Ca{C_{\rm a}}
\def\Cb{C_{\rm b}}
\def\Cavg{C_{\rm avg}}

\def\NB#1{\noindent{\it \textcolor{red}{[#1]}}} 
\def\new#1{{\bf \textcolor{red}{#1}}} 

\authorrunning{L. Fanciullo et al.}

\titlerunning{Polarization and emission-to-extinction ratio}

\maketitle

\section{Introduction}
\label{Section_Intro}

The light of stars often shows a degree of polarization correlated to interstellar extinction, up to a degree of a few percent per $\Av$ magnitude. This phenomenon has long been recognized as the effect of cosmic dust grains aligned with interstellar magnetic field lines \citep{Hall_49, Hiltner_49}. Dust extinguishes starlight, and in the case of non-spherical grains the component of the electric field parallel to a grain's longer axis is more extinguished than the orthogonal one. Furthermore, interstellar dust grains align their shorter axes with the interstellar magnetic field, so they are not generally randomly oriented \citep[\eg,][and refs. therein]{ARAA_review_15}. The overall result is that the dusty, magnetized interstellar medium (ISM) polarizes the starlight that wasn't originally so. The polarization fraction $p$ and the polarization angle $\psi$ of starlight therefore provide information on both interstellar dust and the Galactic magnetic field, or at least the component of it that is parallel to the plane of the sky.

It is mainly the large grains that are aligned \citep[\eg][]{Kim_Martin_95}, and in typical ISM conditions their thermal emission falls mainly in the far-infrared (FIR) and submillimeter (submm) range. This emission is also polarized, since emission is more efficient for the electric field component parallel to the longer axis, and it is an important complement to observations of polarized extinction in the optical and near-infrared \citep[\eg, ][]{Hildebrand_88, Planck_Int_XIX}. It should be noted that, since radiation polarized parallel to grains' longer axis is least intense in extinction and most intense in emission, we expect $\psi$ in the submm to be orthogonal to $\psi$ in the optical.

The main factors that determine the polarization fraction $p$ are: the optical properties of the dust, the alignment efficiency, and the orientation of the magnetic field lines. Polarization can be expressed as \citep{Lee_Draine_85}
$$p = p_0 \, R \, \cos^{2} \gamma$$
where $p_0$ is the maximum possible polarization given the dust properties, the parameter $R$ -- comprised between 0 and 1 -- accounts for the effects of imperfect alignment\footnote{When grains are in the Rayleigh regime -- as is the case for the thermal emission of large dust grains -- the parameter $R$ can be calculated analytically, and it is called the \textit{Rayleigh Reduction Factor} \citep[][p. 328]{Greenberg_68}}, and $\gamma$ is the angle between the magnetic field lines and the plane of the sky. 

The wavelength dependence of polarization in extinction usually follows the so-called \textit{Serkowski curve} \citep{Serkowski}:
$$p(\lambda) = \pmax \cdot {\rm exp}( - K \cdot \ln(\lambda/\lmax)^2)$$
where the polarization has a maximum $\pmax$ at a wavelength $\lmax$, usually falling in the visible; the value of the parameter $K$, tied to the inverse of the FWHM, is usually around unity. Since the polarization efficiency of a grain peaks at $\sim 2\pi$ times its size \citep{Kim_Martin_95}, $\lmax$ traces the typical size of aligned grains: variations in $\lmax$ between lines of sight may indicate a change in grain size distribution, in the dependence of alignment on size, or both \citep[\eg][hereafter \textbf{AP07}]{AP07}. The issue is further complicated by the fact that $\lmax$ also shows some dependence on the magnetic field angle $\gamma$ \citep{Vosh_16}.

Since polarization depends on many factors at once, its interpretation is a degenerate and difficult problem. This is especially true of the dense and complex environments that are molecular clouds, where magnetic field orientation, grain alignment and dust properties are expected to change on small scales. Despite this, studies on dust polarization are often focused on constraining the grain alignment efficiency (\eg~\textbf{AP07}) or the structure of the magnetic field \citep[\eg][]{Planck_Int_XX} without accounting for other variables. One way of confronting the problem is to construct a cloud model that includes dust evolution, magnetic field structure and grain alignment; however, such models are complex and very computationally demanding. The judicious combination of observational data can also provide interesting insights on dust physics while requiring far lighter calculations. 

One example of such labor-saving combinations is the complementary use of extinction data and polarized dust thermal emission. This last one has the dimensions of an intensity and it is usually observed in the far infrared and submillimeter (FIR and submm). All-sky surveys in the submillimetrer such as \Planck\ \citep{Planck_Int_XIX} are opening new possibilities for this kind of multi-wavelength analysis. The idea that extinction and emission combined can be more informative than either of them alone is explored in \cite{Planck_Int_XXI}, which examines the $\Pem/\pv$ ratio\footnote{This ratio is called $R_{\rm P/p}$ in \cite{Planck_Int_XXI}.} in the diffuse ISM, $\Pem$ being the polarized intensity in emission in the \Planck\ 353 GHz channel and $\pv$ being the starlight polarization degree in the $V$ band. This ratio is measured in MJy sr$^{-1}$ and, since $\Pem$ and $\pv$ have (at first approximation) the same dependence on alignment and $\gamma$, it should provide strong constraints on the properties of aligned dust grains. Among the new results made possible by \Planck\ is the determination of average $\Pem/\pv$ in the diffuse ISM: 5.4 MJy sr$^{-1}$, \ie\ about 2.5 times higher than predicted by pre-existing dust models \citep{Planck_Int_XXI}. 

The present paper aims to extend the study of $\Pem/\pv$ to denser environments -- namely, translucent lines of sight in molecular clouds -- using an updated methodology and a dust model optimized for the high $\Pem/\pv$ found by \Planck\ \citep[see][]{Guillet_17}.
The paper is organized as follows: Section \ref{Section_Data} introduces the observational data used, mostly from translucent lines of sight, and the selection that had to be made for the emission/extinction comparison to be meaningful. Section \ref{Section_model} presents the dust model used in our work, especially constructed to reproduce the $\Pem/\pv$ ratio as well as classic dust observables. The dust model was created for the diffuse ISM, while we study translucent clouds where dust evolution may be taking place: the modification that are needed for the model to fit the data hint at the nature of on dust evolution in these areas. The model results are compared to observations in Section \ref{Section_Results}, and the meaning of this comparison is commented in Section \ref{Section_Discussion}. Finally, Section \ref{Section_Conclusions} contains our conclusions and the future perspectives.

\section{Data}
\label{Section_Data}

Our work combines measures of starlight polarization in the near-ultraviolet to near-infrared (NUV to NIR) range, recovered from the published literature, and \Planck\ measurements of total and polarized dust emission at 353 GHz for the same lines of sight. We use a total of 132 objects, which are reduced to 70 after selection (Section \ref{Section_Data_Selection}). 

\subsection{Extinction in the NUV to NIR}
\label{Section_Data_vis}

Most of our data points are from \textbf{AP07} who, in a polarimetric study of the Coalsack nebula, compared their results with data from other clouds [Chamaeleon, Musca, Ophiuchus, R Coronae Australis (RCrA) and Taurus; see Tab.~\ref{tab:Clouds}] taken from pre-existing scientific literature \citep{Covino_97, Arnal_93, Vrba_93, Whittet_92, Whittet_01}.
The \textbf{AP07} study is particularly useful for this purpose because the authors did not employ the Serkowski fit results from the literature, but they used the photometric and polarimetric data therein to conduct their own fits, thus minimizing the systematic effects from different fitting procedures. To increase the statistics we also included data from \cite{Martin_92}, which provides more lines of sight in Ophiuchus, and \cite{Anderson_96}, providing lines of sight not associated, for the most part,\footnote{The star HD 147933 from \cite{Anderson_96} is associated with the Ophiuchus cloud, but this star was eliminated from our sample in the selection process (Section \ref{Section_Data_Selection}).} with the aforementioned clouds. 

\begin{table}\quad
\centering
\begin{tabular}{ l c c c l }
\hline
Cloud & $l$ & $b$ & $D$ (pc) & References \\
\hline
\hline 
Chamaeleon & 297$^\circ$ & -15$^\circ$.5 & 120-150 & 1, 2, \textbf{9} \\
Musca & 301$^\circ$ & -8$^\circ$.0 & 120-150 & 3, \textbf{9} \\
Ophiuchus & 354$^\circ$ & 15$^\circ$.0 & 120 & 1, 4, 5, \textbf{7} \\
R CrA & 0$^\circ$ & -19$^\circ$.5 & 130 & 1, \textbf{8} \\
Taurus & 174$^\circ$ & -14$^\circ$.0 & 140 & 1, 6, \textbf{7} \\
 \, " & 168$^\circ$.5 & -16$^\circ$.5 & " & \, " \\
\hline
\end{tabular}
\caption{
\footnotesize{
Molecular clouds used in this paper, with their approximate position on the sky in Galactic coordinates. Since the Taurus cloud is elongated, stars were sampled from around two different centers. The clouds include most of the lines of sight in this study but not those from \cite{Anderson_96}.\newline
\textsc{References -- } (1) \cite{Whittet_92}; (2) \cite{Covino_97}; (3) \cite{Arnal_93}; (4) \cite{Vrba_93}; (5) \cite{Martin_92}; (6) \cite{Whittet_01}; (7) \cite{Loinard_13}; (8) \cite{Neuh_Forbrich_08}; (9) \cite{Corradi_04}. References in bold provided only the distance estimate and not the polarimetric data. 
}
}
\label{tab:Clouds}
\end{table}

From the literature we obtained the Serkowski polarization parameters $\pmax$ and $\lmax$ for all stars, as well as the polarization angle $\Psiex$. The values of $K$ are not calculated in a consistent fashion in the scientific literature: \textbf{AP07} often use $K = 1.15$ and only fit $K$ as a free parameter if it constitutes a statistically significant improvement, while \cite{Anderson_96} impose that $K$ be a linear function of $\lmax$. For this reason, we chose not to include $K$ in our work. The \textbf{AP07} data retrieved from the \textit{Vizier} online database\footnote{\href{http://vizier.u-strasbg.fr}{http://vizier.u-strasbg.fr}} do not include the uncertainties on $\Psiex$ for Ophiuchus, so we complemented the data with the original article \citep{Vrba_93}: we use the average on $\psi$ in the various bands as the value of $\Psiex$ and their standard deviation as the uncertainty. We excluded the stars with standard deviations greater than 7$^\circ$, which we interpreted as stars where angles in different bands are not compatible. We also obtained the value of the $V$-band polarization $\pv$ for most of the \textbf{AP07} stars, by reading the data directly from the references \citep{Covino_97, Arnal_93, Vrba_93, Whittet_92, Whittet_01}. The data from \cite{Anderson_96} are in the form of (polarized) spectra rather than multi-band photometry; for their lines of sight we took the polarization at $\lambda = 545$ nm as the value for $\pv$. We did not use $\pv$ for the stars in  \cite{Martin_92}, who fit data from multiple sources and thus provide non-unique values for each band. Finally, we obtained the extinction parameters for most of the stars: $\Av$ and $\EBV$ for the \textbf{AP07} stars, $\EBV$ for the \cite{Martin_92} and \cite{Anderson_96} stars. 

\subsection{Emission: \Planck\ and \IRAS\ submm maps}
\label{Section_Data_submm}

Our submm data consists of the \Planck\ 353 GHz (850 $\mu$m) maps for the $I$, $Q$ and $U$ Stokes parameters, from which the polarized intensity $\Pem$ and angle $\Psiem$ were obtained. We did not use any other frequencies because of their lower S/N. For selection purposes we also used the all-sky submm dust opacity maps created by Marc-Antoine Miville-Desch\^enes using \Planck\ and \IRAS\ data \citep{Planck_2013_XI}.

We used the second \Planck\ public data release\footnote{\href{http://irsa.ipac.caltech.edu/data/Planck/release_2/all-sky-maps/}{http://irsa.ipac.caltech.edu/data/Planck/release\_2/all-sky-maps/}}, which consist of HEALPix all-sky maps of 10 quantities: the Stokes parameters $I$, $Q$ and $U$, the number of hits, the variances of the Stokes parameters $II$, $QQ$ and $UU$, and the covariances $IQ$, $IU$ and $QU$. Maps are in NESTED ordering and Galactic coordinates; they have a pixelization $N_{\rm side} = 2048$ for a total of $12 \cdot 2048^2 = 50\,331\,648$ pixels with $1'.7$ side lengths, so that the beam of the instrument (FWHM $\sim 5'$) is well-sampled. The maps are in units of K$_{\rm CMB}$, and are converted to MJy/sr, with a conversion factor of $287.45$ at 353 GHz \citep{Planck_2013_IX}. To obtain the value of $I$, $Q$ and $U$ at the position of each star and increase the S/N, we employed the same technique as \cite{Planck_Int_XXI}: we averaged the values for the Stokes parameters on a Gaussian PSF centered on the star coordinates and with a FWHM of 5', bringing the effective resolution to $\sim7'$. 
In the case of $Q$ and $U$, since we are working on a flat map recovered from a spherical one, we need to account for the fact that the direction of the north changes from pixel to pixel. We do this by rotating the doublet ($Q$, $U$) until it is on the equator in the local reference frame.

With the value of the submillimeter $Q$ and $U$ for all the lines of sight we calculated the polarized intensity in emission, $\Pem = \sqrt{Q^2 + U^2}$, and the polarization angle $\Psiem = \frac{1}{2} \arctan{(U,Q)}$ using the HEALPix angle convention (where the relative signs of $Q$ and $U$ are inverted with respect with the IAU convention). Being a quadratic function of measures with finite noise, $\Pem$ has a positive bias and it was debiased with the conventional formula \citep{Wardle_Kronberg_74}: $P_{\rm deb} = \sqrt{P_{\rm bias}^2 - \sigma_P^2}$. We did not apply the more recent debiasing method \citep[\eg][]{Plasz_13, Montier_15} because, after the smoothing, the environments we are studying have a high signal-to-noise ratio, and therefore a low bias. There was no need to apply CMB and CIB corrections, which are negligible at this wavelength and for our datset. 

For each star we also calculated the polarization angle dispersion function $S$ \citep{Planck_Int_XIX}, a tracer of disorder in polarization and therefore in the magnetic field orientation. The $(I,Q,U)$ triplet was again smoothed to increase S/N, using a Gaussian PSF with 5' FWHM and bringing the maps to a $\sim7'$ resolution. The maps thus obtained are oversampled (4 pixels per beam), so we also degrade the pixelization of the $Q$ and $U$ maps to $N_{\rm side} = 1024$ to get closer to the Nyquist criterion. The dispersion function $S$ is then computed for the pixel containing the star, with a lag $\delta$ = 5'.

\subsection{Selection}
\label{Section_Data_Selection}

\begin{figure}
\includegraphics[width=\hsize]{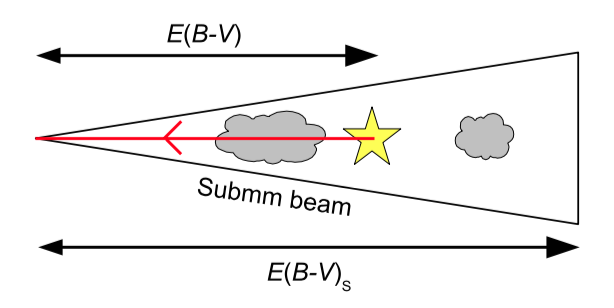}
\caption{
\footnotesize{
Illustration of line-of-sight and beam effects when comparing extinction and emission, from \cite{Planck_Int_XXI}. The measured extinction is \EBV, the extinction obtained from conversion of submillimeter emission is \EBV$_{\rm S}$. 
}
}
\label{Contamination_Pic}
\end{figure}

Since this paper compares different phenomena (dust extinction and thermal emission) observed at different wavelengths (NUV-to-NIR vs. submm), we need to ensure that the comparison is meaningful \citep[see \eg][]{Planck_Int_XXI}. This amounts to making sure that first, we are observing the same type of grain at the two wavelengths; second, the two wavelengths probe the same volumes of ISM. 

The first condition is met, in first approximation, as a consequence of dust physics: only large grains  contribute to polarization, in extinction and in emission. Large grains are also the main contributor to submm emission as well as visual and IR extinction (small grains are important in the UV). It should be kept in mind, however, that polarization and overall extinction do not necessarily trace the same grains: a population of large grains that are spherical or unaligned would contribute to extinction and emission, but not polarization. 

The second condition is not trivial. Extinction measured on background stars -- as is the case of the data described in section \ref{Section_Data_vis} -- only probes the matter in front of the star itself; emission has no such limitation, especially in the submm where the ISM is optically thin (see Fig.~\ref{Contamination_Pic}). In presence of a background to the star, the total intensity $I$ measures systematically more dust than what is observed in the optical. The effect of background on polarization is more complex, and we will detail it in Section \ref{Section_Data_depol}. To compare extinction and emission, therefore, we need to discard those lines of sight that have significant dust emission from behind the star. As shown in \cite{Planck_Int_XXI}, this selection can be based on three criteria:
\begin{description}
\item[\bf{Galactic latitude:}] All stars close to the Galactic plane are very likely to have significant background; so we only keep stars stars with Galactic latitude $\vert b \vert \ge 2^\circ$. This forces us to exclude from the study some well-studied clouds, such as the Coalsack Nebula (\textbf{AP07}), located on the Galactic plane.
\item[\bf{Polarization angle:}] Dust polarization in extinction should be orthogonal to that in emission.  In lines of sight where the angles $\Psiex$ and $\Psiem$ are not orthogonal, extinction and emission do not come from the same dust; we exclude from our sample such lines of sight with a tolerance $3\sigma$ or $10^\circ$, whichever is smaller. For stars in \cite{Anderson_96}, whose $\Psiex$ are given without uncertainties, we assume $\sPsiex = 0$. Since \Planck\ angle uncertainties are usually larger than $V$ band angle uncertainties, this should not make a large difference.
\item[\bf{Column density:}] The dust submm optical depth $\tem$ can be converted to an expected $\Av$ or $\EBV$ and compared to the actual extinction measured; lines of sight where the $\tem$-derived extinction shows an excess have significant background. We use for this the empirical conversion factor $\EBV/\tem = 1.49\,10^4$ obtained by \cite{Planck_2013_XI} for the diffuse ISM. In molecular clouds, however, $\tem/\EBV$ is known to increase by a factor $\sim 2-3$ compared to the diffuse ISM \citep[\eg][]{Stepnik_03, Ysard_13, Planck_2013_XI}. We decided therefore to relax this condition and keep all lines of sight where the $\tem$-derived $\EBV$ is less than 3 times the measured value.
\end{description}

While none of these selection procedures intrinsically exclude all of the contaminated sightlines, when combined, and used together with the selection already operated by \textbf{AP07}, they are more robust. 

We did some additional selection to improve the data quality. We only kept those stars for which we had a S/N greater than 3 in $\Pem$ and greater than 5 in $\lmax$. For a few of the stars in \textbf{AP07} the quality of the Serkowski fit was low and the the Serkowski parameters were not an adequate representation of the polarization curve; we recovered the observational data from  \cite{Covino_97, Arnal_93, Vrba_93, Whittet_92, Whittet_01} and excluded those stars that do not follow Serkowski. Finally, we excluded those stars that according to \cite{Anderson_96} are likely to have intrinsic polarization. The combined selection left us with the values of $\lmax$, $\pmax$, $\EBV$, $\Iem$ and $\Pem$ for 70 lines of sight, 56 of which also have information on $\Av$ and $\pv$.

\subsection{Line-of-sight and beam depolarization}
\label{Section_Data_depol}

The magnetic field in the ISM has a non-negligible disordered (or ``meandering'') component that introduces a confounding variable called ``depolarization''. When in an observation there is confusion between polarized sources with different orientation angles, it is possible for orthogonal components of polarization to cancel each other out, so the overall polarization observed may be lower than that of each source taken separately. Depolarization may occur if the interstellar magnetic field changes orientation along a line of sight (line-of-sight depolarization), or if an instrument has a finite observational beam and the magnetic field changes orientation on scales smaller than said beam (beam depolarization). In most polarization studies, the two effects are put together under the name of ``beam depolarization'' or simply ``depolarization''. However, since the two types of depolarization have different effects on the extinction and the emission, we will treat them separately in the present paper. 

The line-of-sight depolarization, at first approximation, has the same effect on extinction and emission if they probe the same ISM. Complications arise if there is significant emission from the background to the star: if the magnetic field orientation is very different in the foreground and in the background, depolarization in emission may be very different from that in extinction. This would give unreliable measurements of, \eg, the ratio $\Pem/\pv$. The selection described in Section \ref{Section_Data_Selection}, if effective, should ensure that the line-of-sight depolarization affects extinction and emission in the same way. We remark that having a uniform magnetic field orientation on the line of sight is not equivalent to having no background emission: since $\Pem$ is additive, in this case we would observe an excess of polarization in emission as compared to extinction, and overestimate $\Pem/\pv$.

The beam depolarization affects observations that have finite beam size. This is usually the case at FIR and submm wavelengths: the \Planck\ beams measure 5' or more. Extinction observations on stars, on the other hand, are pointlike and suffer no beam effects, so that beam depolarization affects only polarized emission. Unlike line-of-sight depolarization, beam effects are unaffected by our selection criteria. However, the amount of beam depolarization can be estimated from observational data, such as the function $S$ (Section \ref{Section_Data_submm}) that measures field disorder.

\subsection{Final observables}
\label{Section_Data_final}

\begin{figure}
\includegraphics[width=\hsize]{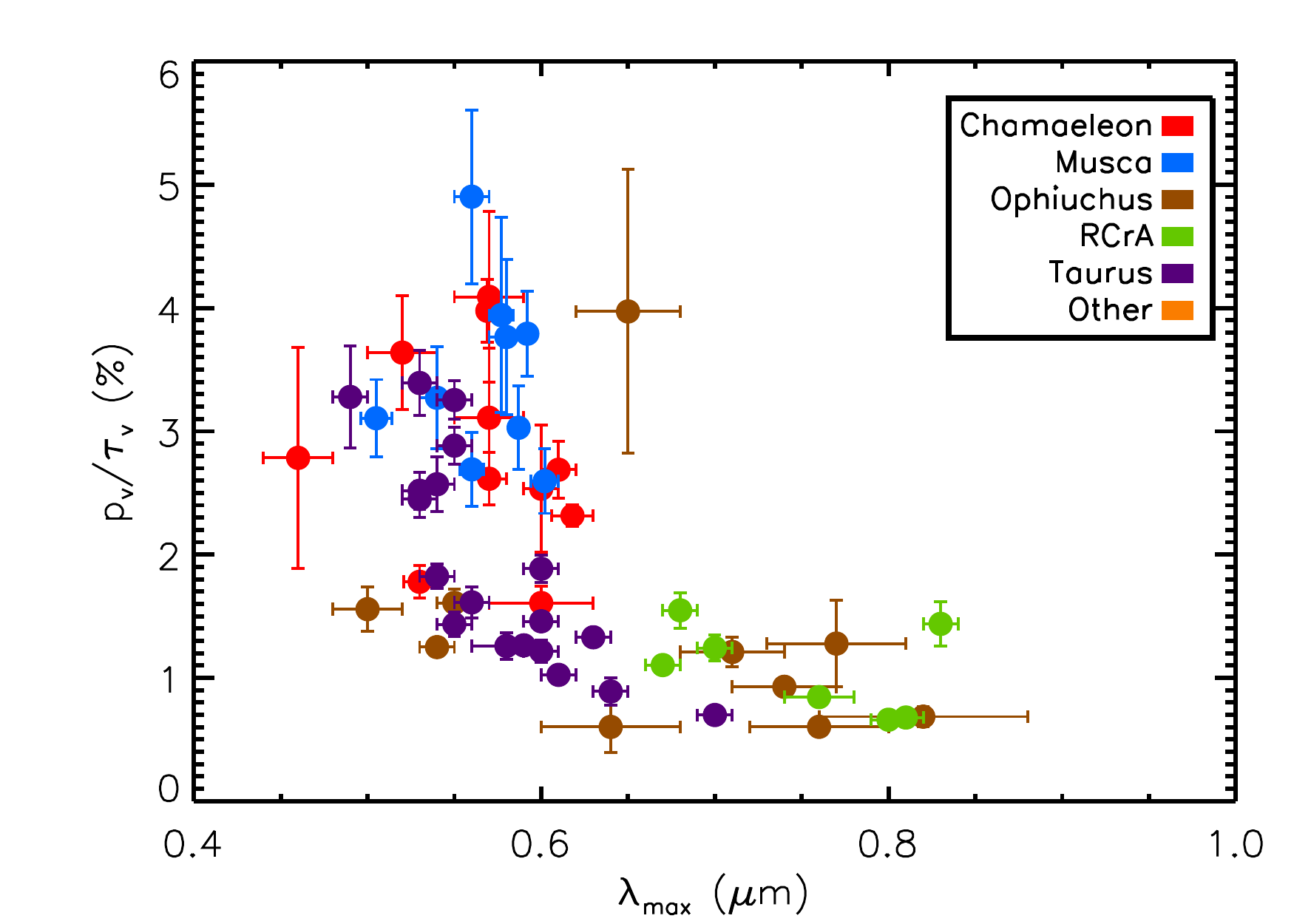}
\includegraphics[width=\hsize]{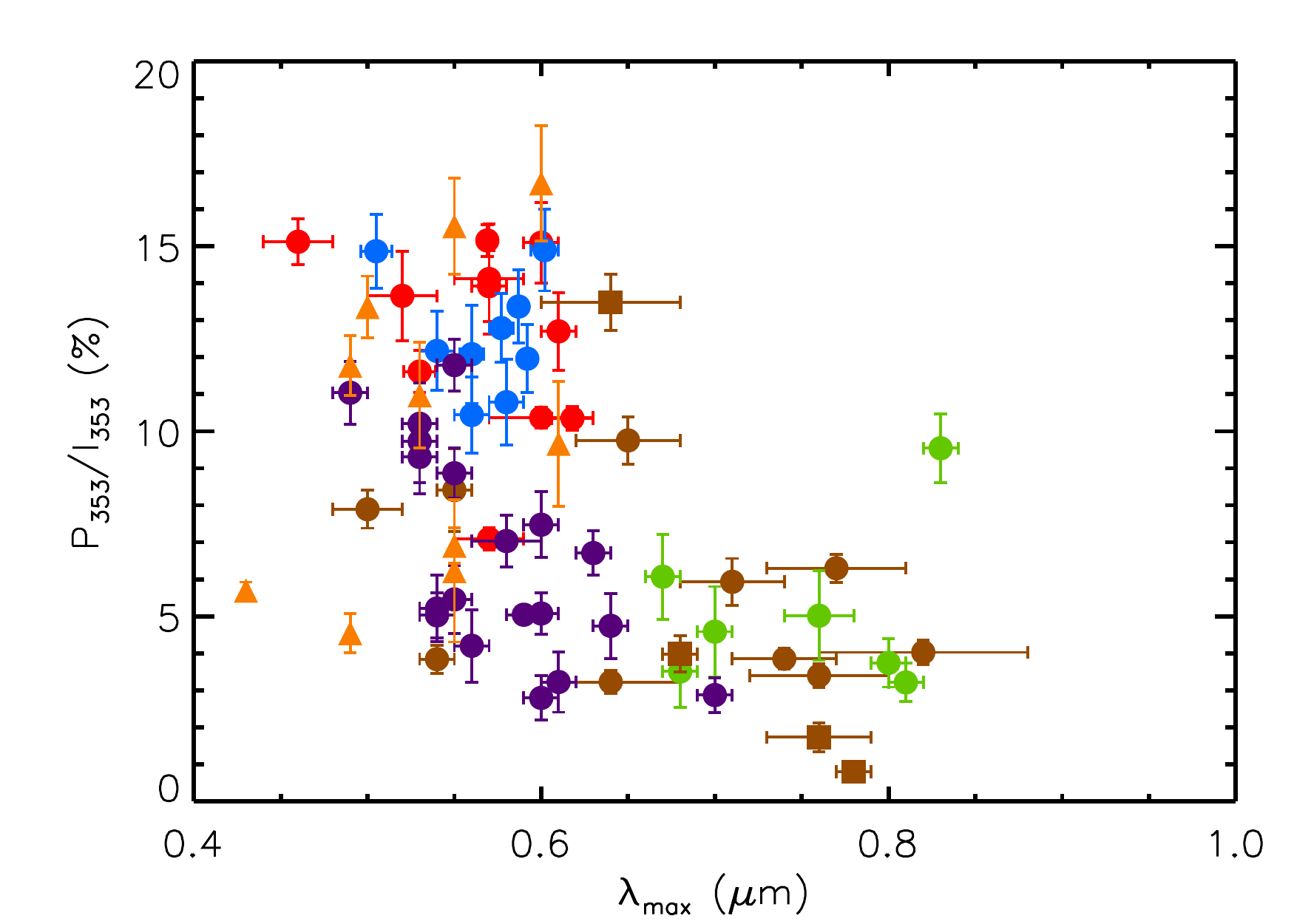}
\includegraphics[width=\hsize]{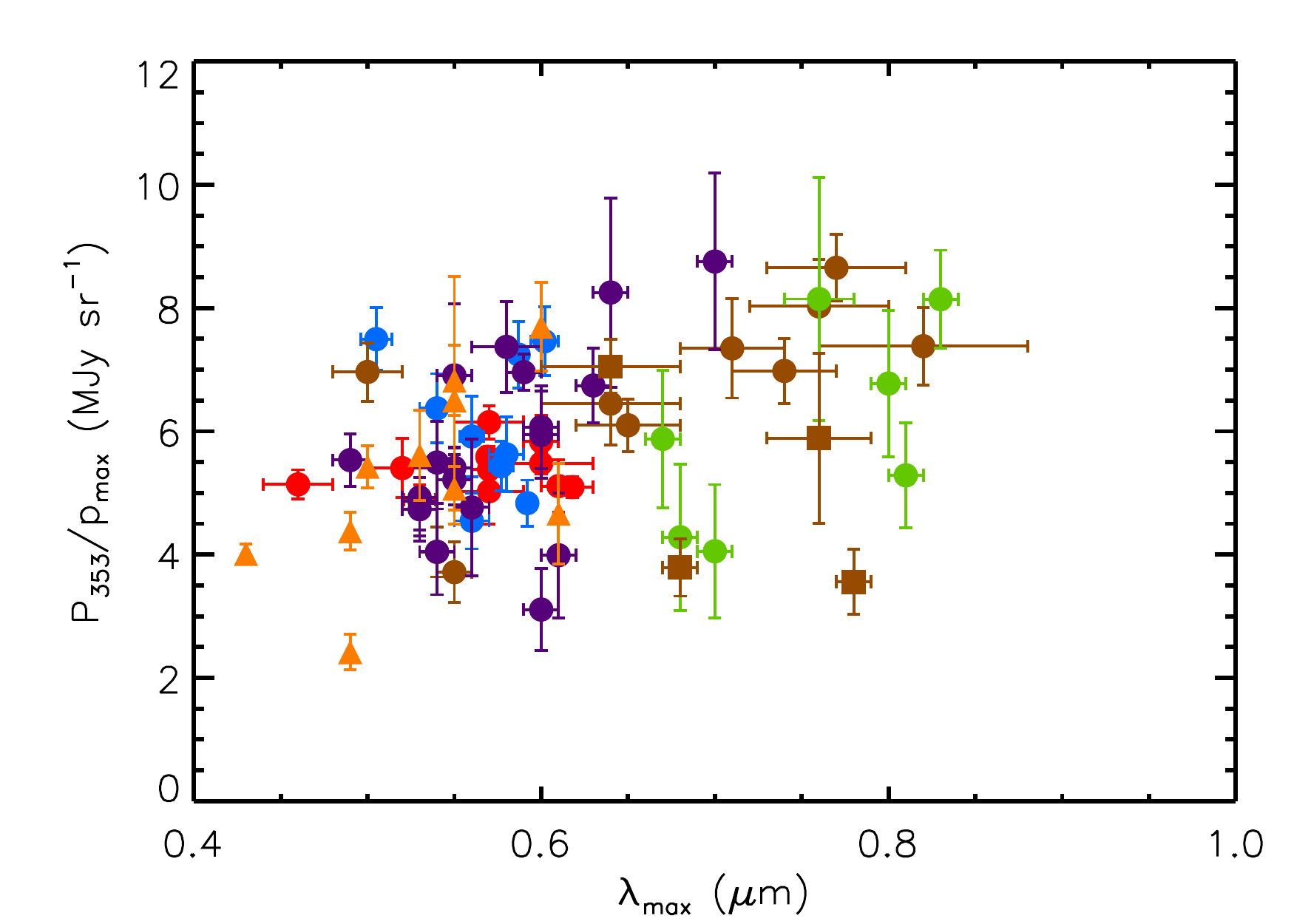}
\caption{
\footnotesize{
The observational data to reproduce. Different colors indicate different clouds, different symbols indicate different references (circles: \textbf{AP07}; triangles: \cite{Anderson_96}, squares: \cite{Martin_92}). \textit{Top}: the ``classical'' $\pv/\tau$ vs. $\lmax$ plot. This panel does not contain the stars without $\Av$ and $\pv$ measurements. \textit{Center}: comparison of $\lmax$ in extinction with the polarization $\Pem/\Iem$ in emission. \textit{Bottom}: the ratio of polarization $\Pem/\pmax$ as a function of $\lmax$. 
}
}
\label{Obs_Lmax_correlations}
\end{figure}

Our observational data, to be compared to a dust model, is plotted in Fig.~\ref{Obs_Lmax_correlations}. The top panel shows the normalized polarization in extinction, $\pv/\tv$, as a function of $\lmax$. The two quantities have a clear negative correlation; we also see that the values for polarization are very widely scattered, their upper limit marking an ``envelope'' -- as is typical of polarized observations -- the shape of which may be partly determined by line-of-sight depolarization \citep[\eg][]{Planck_Int_XX}. A very similar behavior can be seen in the submm polarization fraction $\Pem/\Iem$ as a function of $\lmax$, shown in the central panel. The bottom panel of Fig.~\ref{Obs_Lmax_correlations} shows a different observable: $\Pem/\pmax$. This quantity, like the ratio $\Pem/\pv$ used in \cite{Planck_Int_XXI}, is meant to trace the optical properties of grains by normalizing out the effects of alignment and magnetic field orientation affecting both emission and extinction. We decided to use $\pmax$ in the construction of this ratio, as opposed to $\pv$, to avoid introducing spurious correlations: many of our stars have high values of $\lmax$ ($> 0.6 \, \mu$m), and $\pv$ and $\lmax$ are going to be negatively correlated in that range. The bottom panel of Fig.~\ref{Obs_Lmax_correlations} shows that, even if the scatter in $\Pem/\pmax$ is quite large, it is still small compared to the one observed in $\pv/\tv$ and $\Pem/\Iem$; also the dependence of  $\Pem/\pmax$ on $\lmax$ is much less pronounced. This is consistent with our expectations that this quantity be nearly independent of alignment and magnetic field orientation.

\section{Model: DUSTEM with polarization}
\label{Section_model}

\begin{table*}\quad
\centering
\begin{tabular}{ l c c c c c c c }
\hline
Population & mass (per H) & depletion (ppm) & $\amax \, (\mu{\rm m})$ & $\alpha$ & $\athresh \, (\mu{\rm m})$ & $p_{\rm stiff}$ & $f_{\rm max}$ \\
\hline
\hline 
PAHs & $7.10 \times 10^{-4}$ & 59 & -- & -- & -- & -- & -- \\
Carbon BGs & $1.32 \times 10^{-3}$ & 110 & 0.07 & - 4.14 & -- & -- & -- \\
Silicate BGs & $6.52 \times 10^{-3}$ & 37.9 & 0.52 & - 3.32 & 0.108 & 0.27 & 1.00 \\
\hline
\end{tabular}
\caption{
\footnotesize{
Standard version of the dust model used in the present article \citep[model A from][]{Guillet_17}. ``Depletion'' in the case of silicates refers to Si, Mg and Fe. The interstellar radiation field used in the model is the \cite{Mathis_83} spectrum for the Solar neighborhood. 
}
}
\label{tab:Model}
\end{table*}

\begin{figure}
\includegraphics[width=\hsize]{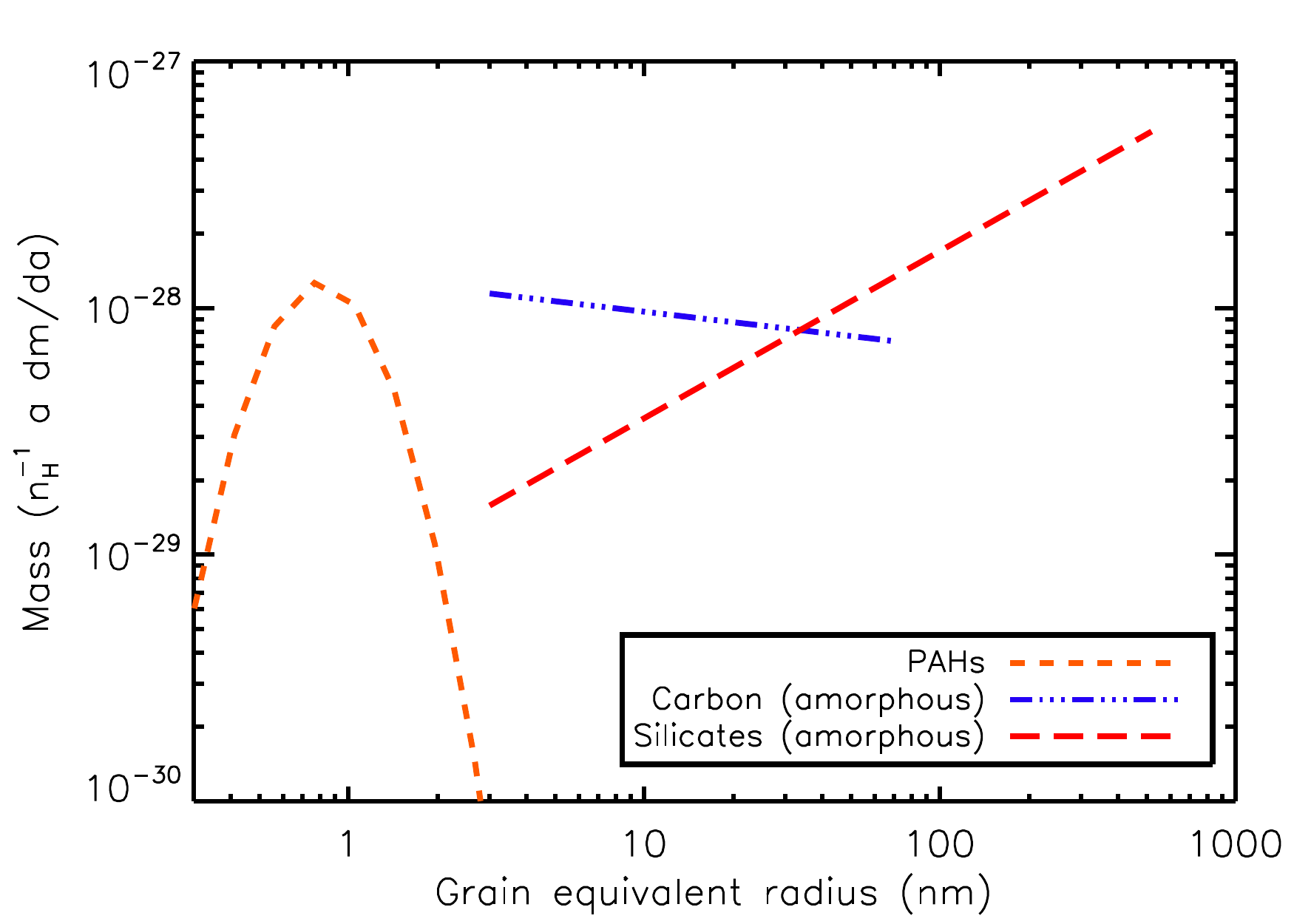}
\caption{
\footnotesize{
Size distribution for the model used in the present paper \citep[model A from][]{Guillet_17}. 
}
}
\label{model_sdist}
\end{figure}

The dust model we use should ideally have the three following characteristics: it should predict polarization in both extinction and emission, it should be compatible with the latest results from the \Planck\ mission -- especially the $\Pem/\pmax$ ratio, which is underpredicted by pre-\Planck\ models -- and it should allow to modify the dust properties to simulate dust evolution. Unfortunately, while models that allow detailed dust evolution exist, to the authors' knowledge they either do not predict polarization \citep[\eg][]{Jones_13} or they are calibrated on extinction alone and cannot be expected to reproduce the correct $\Pem/\pmax$ ratio \citep[\eg][]{Hirashita_Vosh_14}. We decided to instead use a model optimized for fitting the latest \Planck\ data, at the cost of a simplified treatment of dust evolution where only grain size is accounted for.

We adopt the dust model recently developed by \cite{Guillet_17} and called ``Model A'', a modified version of \cite{Compiegne_11}. The computation is done with the DustEM Fortran numerical tool\footnote{\href{https://www.ias.u-psud.fr/DUSTEM/}{https://www.ias.u-psud.fr/DUSTEM/}} and its IDL wrapper.\footnote{\href{http://dustemwrap.irap.omp.eu/}{http://dustemwrap.irap.omp.eu/}} The model populations and size distributions are chosen to minimize the number of free parameters; the parameters themselves are calculated by fitting the observables typical of the low-latitude diffuse ISM $(\vert b \vert < 30^\circ)$: the extinction curve, the polarization in extinction up to $4 \, \mu$m, and the emission and polarization SED updated with \Planck\ results. The model therefore reproduces the average observations for the diffuse ISM, including ratios such as $\Iem/\Av$ and $\Rpp$. 

The model includes three grain types (see Tab. \ref{tab:Model} and Fig.~\ref{model_sdist}): a population of neutral PAHs with a lognormal size distribution, plus two populations of big grains (amorphous carbon and silicates, respectively) distributed as power laws: $dn/da = a^\alpha$. We are mainly interested in observables where the big grain contribution is dominant, so the model, unlike \cite{Compiegne_11}, has no separate population for very small carbonaceous grains: the very small grains are included in the amorphous carbon population, which is why the power law for carbons is weighted towards small sizes. Large grains are prolate spheroids with an axial ratio of 3 \citep[oblate grains of the same axial ratio cannot reproduce the high $\Pem/\Iem$ observed by \Planck:][]{Guillet_17}. The neutral PAHs and the amorphous carbon grains have the same compositions as their counterparts in \cite{Compiegne_11}; the silicate grains have the same composition as \cite{Weingartner_Draine_01}, with added porosity: 20\% of their volume consists of vacuum inclusions. The porosity of the silicate grains is essential in increasing their $\Rpp$ ratio to the value observed by \Planck.

\begin{figure}
\includegraphics[width=\hsize]{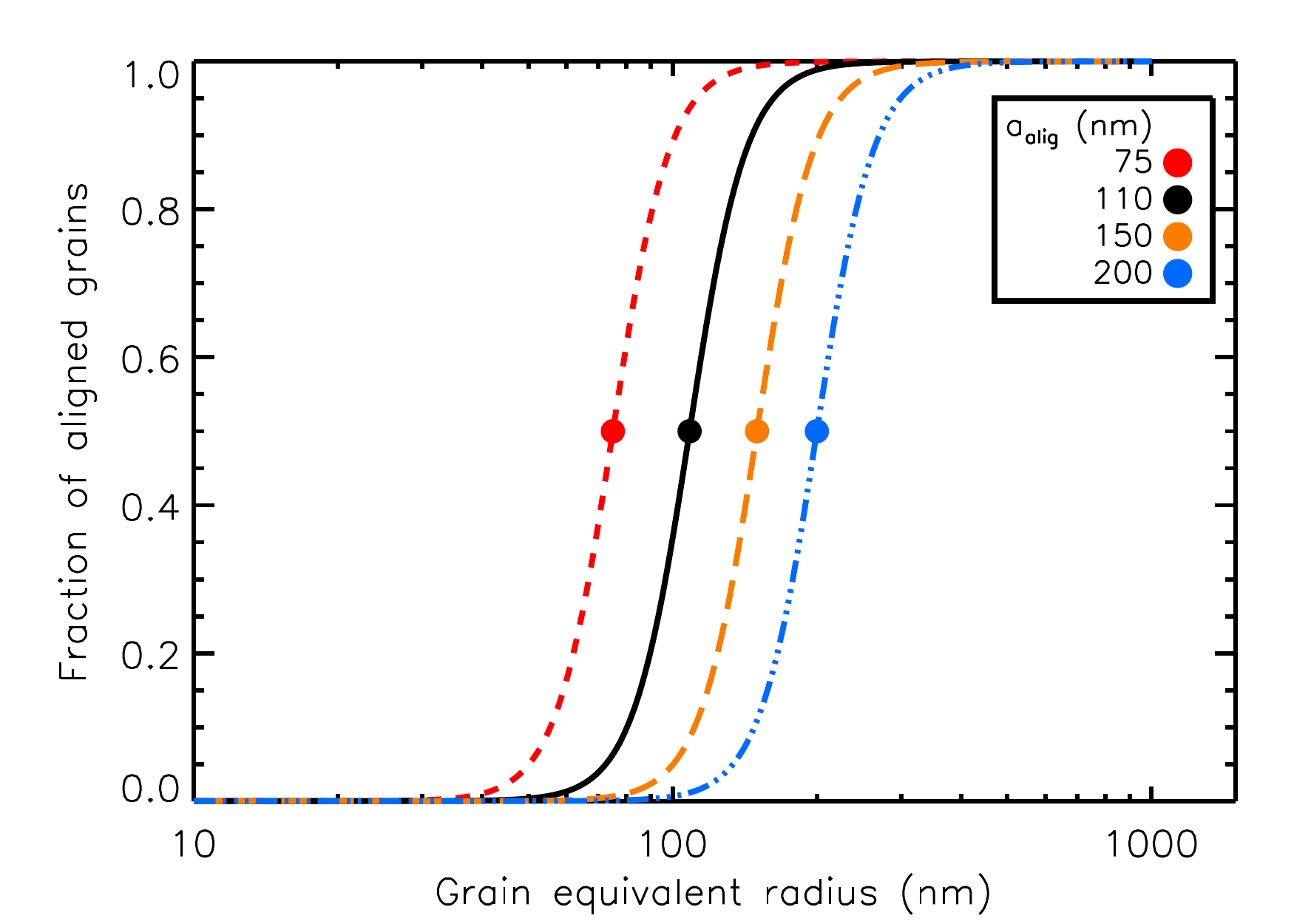}
\includegraphics[width=\hsize]{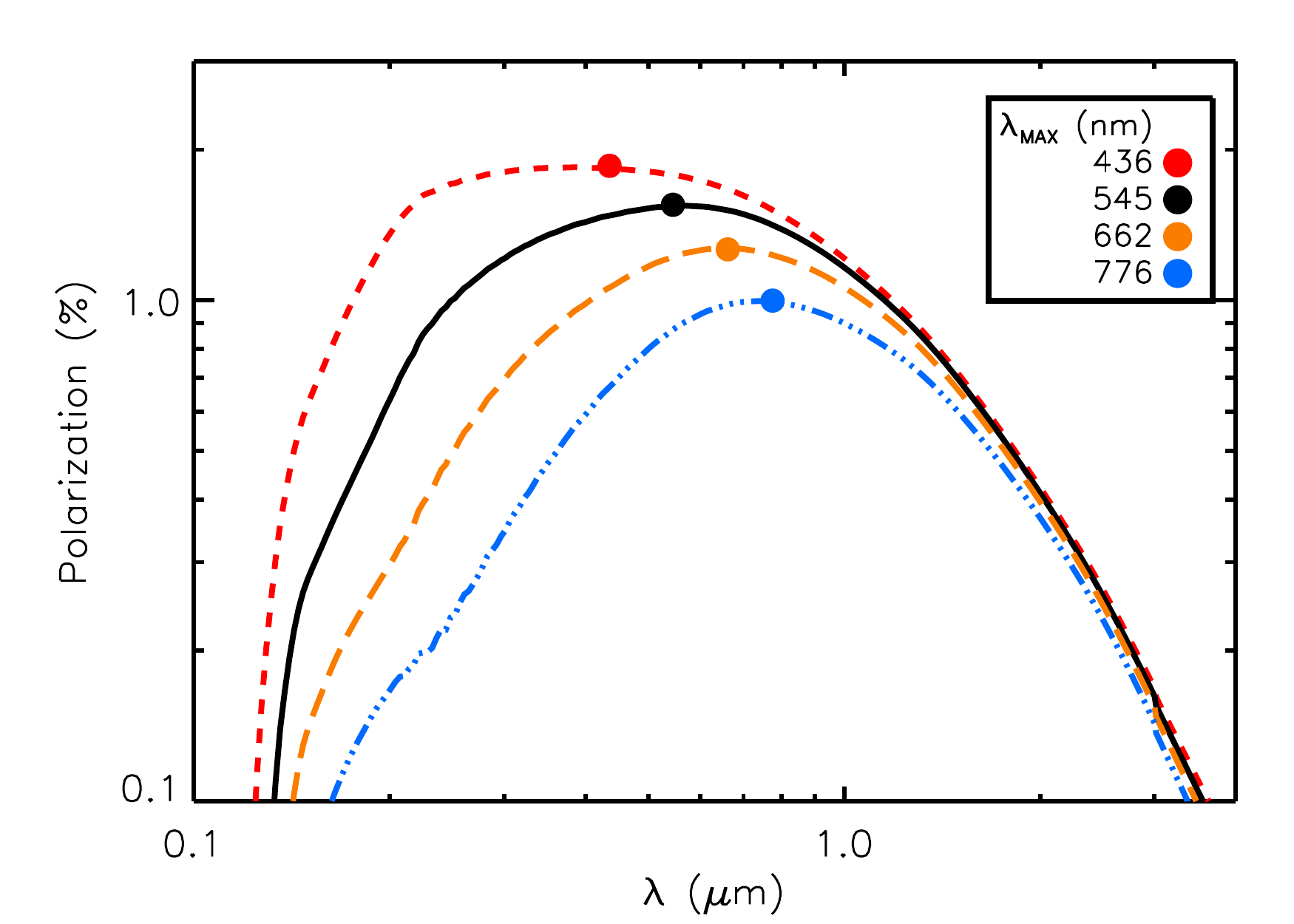}
\caption{
\footnotesize{
\textit{Top}: grain alignment function for different values of $\athresh$. \textit{Bottom}: corresponding polarization curves in extinction, normalized to $\Nh = 10^{21} \, {\rm cm}^{-2}$. Dots indicate the value of $\athresh$ (top) and that of $\lmax$ from free-$K$ fits (bottom). 
}
}
\label{athresh_effects}
\end{figure}

In the model silicate grains are aligned according to the phenomenological alignment function provided by DustEM:
\begin{equation}
\label{eq_ch4_falign}
f(a) = \frac{1}{2} f_{\rm max} \, \left( 1 + \tanh{\left( \frac{\ln (a/\athresh)}{p_{\rm stiff}} \right)}\right)
\end{equation}
where $f(a)$ is the fraction of grains aligned as a function of the equivalent radius\footnote{For non-spherical grains, the equivalent radius is the radius of a sphere of corresponding volume.} $a$, $f_{\rm max}$ is the maximum alignment efficiency, $\athresh$ is the size threshold for grain alignment and the parameter $p_{\rm stiff}$ denotes the width of the transition. This alignment function is designed to increase monotonically with size, since small grains are generally unaligned. This parametric function is not designed to test any particular alignment process, and while its shape resembles the typical result of the radiative torque model \citep{Lazarian_Hoang_07}, it is also compatible with magnetic alignment for grains with superparamagnetic inclusions  \citep{Mathis_86, Vosh_16}. The variation of the alignment function for varying $\athresh$ and its effects on the polarization curve in extinction, are shown in Fig.~\ref{athresh_effects}.

In addition to causing polarization, grain alignment affects dust extinction and emission as well. The resulting correction is very small and it is generally ignored in dust models; nonetheless, DustEM provides the option of including alignment effects in extinction and emission. In the present paper, we chose to ignore these effects to allow a more direct comparison to the results of \cite{Guillet_17}, where minor alignment effects were ignored as well. 

\subsection{Model results: fitting}
\label{section_model_fitting}

DustEM provides the full extinction curve and dust emission for the model, including polarization; most of the parameters introduced in Section \ref{Section_Data_final} have to be obtained from a fit to the DustEM output.

We take the extinction curve interpolated at 550 nm to be $\Av$ of the model, and the thermal emission at 353 GHz -- plus a color correction to account for the spectral response of the corresponding \Planck\ band -- as the value $\Iem$ of the SED. The same interpolation, operated on the polarized extinction and emission, gives us the model prediction for $\pv$ and $\Pem$. The $\pmax$, $\lmax$ and $K$ of the model are calculated by interpolating  the model polarized extinction at the effective central wavelengths of the $UBVRIJH$ photometric bands (Tab.~\ref{tab:UBVRIJH}) and fitting a Serkowski function to the synthetic observations thus obtained, keeping $K$ as a free parameter. 

\begin{table}\quad
\centering
\begin{tabular}{ c c }
\hline
Band & Effective $\lambda$ ($\mu$m) \\
\hline
\hline
U & 0.36 \\
B & 0.44 \\
V & 0.55 \\
R & 0.65 \\
I & 0.80 \\
J & 1.25 \\
H & 1.60 \\
\hline
\end{tabular}
\caption{
\footnotesize{
The effective wavelengths used to interpolate the model polarization curve and produce synthetic $UBVRIJH$ observations for the Serkowski fit. Adapted from \cite{Whittet_92, Whittet_01, Bessell_Murphy_12}.
}
}
\label{tab:UBVRIJH}
\end{table}

We also fitted the model SED with a modified blackbody: $I_\lambda = B_\lambda(T) \cdot \tau_0 \cdot (\lambda/\lambda_0)^{-\beta} $. The fit was performed on the model emission interpolated and color-corrected at the wavelengths of the \Planck\ HFI bands (350, 550, 850, 1380 and 2100 $\mu$m) as well as the \IRAS\ 100~$\mu$m band. Since emission at those wavelengths is dominated by big grains, integrating the modified blackbody over wavelength provides the radiance $\mathcal{R}$, or emitted power, of the big grain populations \citep[see][]{Planck_2013_XI}.

\subsection{Model variations}
\label{Section_model_var}

The model so far described has been developed to fit the average observables in the diffuse ISM. 
Inside molecular clouds, however, evolution alters the properties of dust considerably, the main alteration being grain growth due to accretion and coagulation \citep{Boulanger_90, Stepnik_03, Kohler_12}. Translucent lines of sight, such as those studied in this paper, are typically at the onset of such evolution: for instance, \cite{Stepnik_03} and \cite{Ysard_13} find that coagulation takes place where $\Av$ is greater than 2 or 3. This raises the question of whether our sample can be explained by a dust model designed for the diffuse ISM. To answer this question, we studied how the model output is affected by grain alignment efficiency, magnetic field orientation and grain size. Comparing these results to the observations will inform us whether the data can be explained by the variation of alignment and field orientation alone, using the same dust as in the diffuse ISM, or whether dust growth is necessary, and to what extent. The modifications to our model are purely phenomenological, not based on simulations of grain growth, dust alignment or magnetic field structure; however, they are useful for estimating the variations that physical models will have to reproduce.

\subsubsection{Variations of the alignment function}
\label{Section_model_var_alig}

Loss of grain alignment inside molecular clouds is sometimes invoked to explain the decrease in polarization efficiency observed at high $\Av$ \citep[\eg][and refs. therein]{ARAA_review_15}. This weakening of polarization is also in qualitative agreement with some alignment theories, \eg\ the radiative torque model that predicts that only the largest grains are aligned inside molecular clouds.

The alignment efficiency in DustEM is a function of the three parameters $\athresh$, $p_{\rm stiff}$ and $f_{\rm max}$, as shown in Eq.~\ref{eq_ch4_falign}. We simulated different alignment efficiencies by running the model with different values of $\athresh$, keeping the same $p_{\rm stiff}$ for simplicity. We did not change $f_{\rm max}$ (equal to 1 in our model) because it has the exact same effects on $\pv$ and $\Pem$, and it could have no effect on $\Rpp$; the parameter $\athresh$, on the other hand, affects $\pv$ and $\Pem$ similarly but not identically, since polarization cross-section has different size-dependent behavior in the visible (where scattering is dominant) and in the submillimeter.

Fig.~\ref{athresh_effects} shows how $f(a)$ changes as a function of $\athresh$ and the effect this has on the polarization curve; higher $\athresh$ corresponds to lower $\pmax$, since fewer grains are aligned, and higher $\lmax$, since the average aligned grain is larger.

\subsubsection{Variations in magnetic field orientation}

An ordered magnetic field forming an angle $\gamma$ with the plane of the sky introduces a factor  $\cos^{2} \gamma$ in the polarized intensity, as shown in Section \ref{Section_Intro}. Dust polarization models would greatly benefit from measures of the angle $\gamma$; unfortunately, dust only traces the magnetic field component parallel to the plane of the sky, so this information is not usually available. The angle $\gamma$ is therefore another variable parameter in our model: we ran the model for $\gamma = 0^\circ$, $30^\circ$, $45^\circ$ and $60^\circ$. 

The Galactic magnetic field is actually the sum of an ordered component and a disordered, or meandering, one: this latter component causes the phenomena known as line-of-sight and beam depolarization, as explained in Section \ref{Section_Data_depol}. Our model does not include a disordered magnetic field component and therefore it cannot predict depolarization; however, 
the polarization angle dispersion $S$ (Section~\ref{Section_Data_submm}) can be used as a measure of field disorder \citep[see \eg][which compare the observed $S$ with MHD simulations]{Planck_Int_XX}. Using this we were able to assess some of the effects of the disordered magnetic field, as shown in Section \ref{Section_Results_Alig+Field}.

\subsubsection{Variations of grain size distribution}
\label{Section_model_var_sdist}

Gas accretion on grain surfaces \citep[\eg][]{Jones_13} and formation of aggregates are known to increase grain sizes inside molecular clouds. This growth is supported by theoretical studies \citep{Kohler_12, Hirashita_Vosh_14} and it is consistent with observed phenomena such as the flattening of extinction curves in dense environments \citep[\eg][]{Fitzpatrick_99, Weingartner_Draine_01} and the coreshine observed in the NIR \citep{Pagani_10}. 

As already mentioned, there are for now no dust models that treat dust evolution realistically while reproducing the emission-to-extinction polarization ratio revealed by \Planck. We therefore opted for a model that is compatible with \Planck\ data \citep[model A from][]{Guillet_17} at the cost of a simplified treatment of dust evolution. Specifically, we will focus on a single aspect affected by dust evolution: the size distribution of grains. In our model, the size distribution for big grains is a power law defined by three parameters: the minimum and maximum grain sizes $a_{\rm min}$ and $\amax$, and the power law index $\alpha$.

In the case of silicates most of the mass is in the large grains (see Fig.~\ref{model_sdist}), meaning that the size distribution is most sensitive to $\amax$; furthermore, none of the observables we use are in the UV where the contribution of small grains is important. Therefore, we decided to vary the $\amax$ of silicates between 350 nm and 1 $\mu$m (the standard value is $\sim 500$ nm) and keep $\alpha$ and $a_{\rm min}$ fixed. Although we chose to fix $\alpha$ mainly as a matter of convenience, we note that this is consistent with the model of grain growth by \cite{Hirashita_Vosh_14}, where the slope of the size distribution does not change much during evolution, and the largest variation is in the upper size cutoff.

The case of carbon grains is different, as their distribution is weighted towards small sizes: the mass available for large grains is now also dependent on $a_{\rm min}$ and on the amount of PAHs, so a realistic model becomes a necessity. For this reason we only varied the size distribution of silicates while leaving that of carbon grains fixed. While this choice does not give realistic results for the variation of extinction and emission with size distribution, it still allows us to predict the dust polarization, which in our model depends on silicates alone.

\subsubsection{Multi-parameter study: Monte-Carlo}
\label{Section_Model_MC}

The phenomena described in sections \ref{Section_model_var_alig} to \ref{Section_model_var_sdist} are all expected to occur in molecular clouds, so that variations of a single model parameter at a time are not realistic, even if studying their effect can be instructive. We decided to use a Monte-Carlo simulation to explore the effects of simultaneous variations of many parameters. As explained in the previous section, we can only vary the size distribution of silicates, which gives realistic results for polarization but not for unpolarized observables; therefore our Monte-Carlo results can only be compared to polarization observables. The model was run one thousand times, and the values of $\athresh$ and $\amax$ for silicates were uniformly distributed within the ranges 
\begin{center}
$50 \, {\rm nm} < \athresh < 300 \, {\rm nm} $\\
$350 \, {\rm nm} < \amax < 1000 \, {\rm nm} $\\
\end{center}
The operation was repeated for four values of $\gamma$ ($0^\circ$, $30^\circ$, $45^\circ$ and $60^\circ$), bringing the simulations to a total of 4000.

We found that not all combinations of $\athresh$, $\amax$ and $\gamma$ in the ranges chosen are realistic: the synthetic observables obtained for some such combinations have values that are never observed. We set out to find a realistic range of parameters by imposing that model results have a range as close as possible to that of actual observations \citep[\eg, that $0.35 < \lmax < 0.8 \, \mu$m and $0.5 < K < 1.5$ as per][]{Vosh_Hiro_14}. Restricting $\athresh$ and $\amax$ to the following ranges eliminates most of the unrealistic values for $\lmax$ and $K$:
\begin{center}
$75 \, {\rm nm} < \athresh < 150 \, {\rm nm} $\\
$350 \, {\rm nm} < \amax < 800 \, {\rm nm} $\\
\end{center}
while the same four values for $\gamma$ are kept. This selection left us with 844 Monte-Carlo iterations.

\section{Results}
\label{Section_Results}

\subsection{Alignment efficiency and magnetic field orientation}
\label{Section_Results_Alig+Field}

\begin{figure}
\includegraphics[width=\hsize]{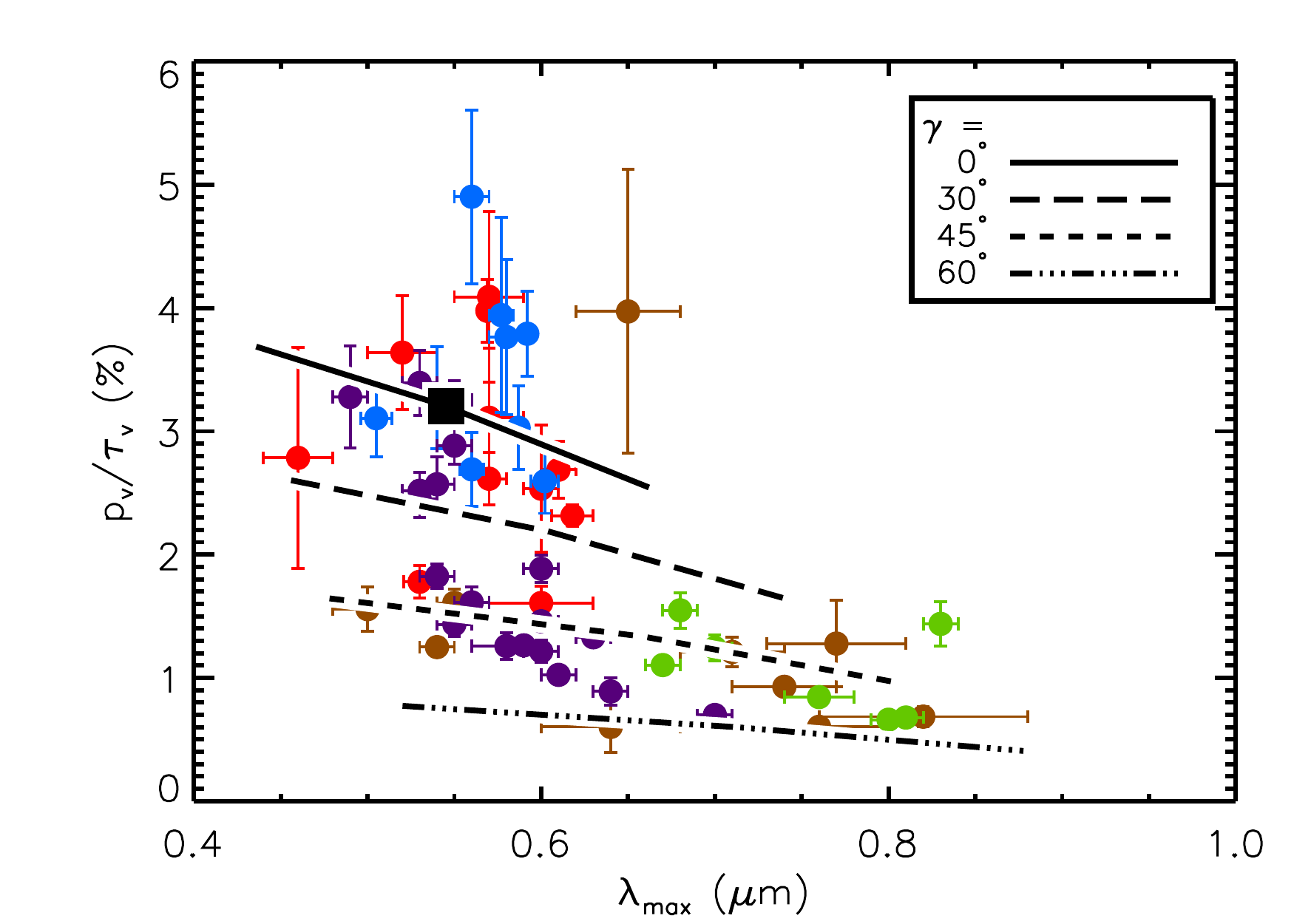}
\includegraphics[width=\hsize]{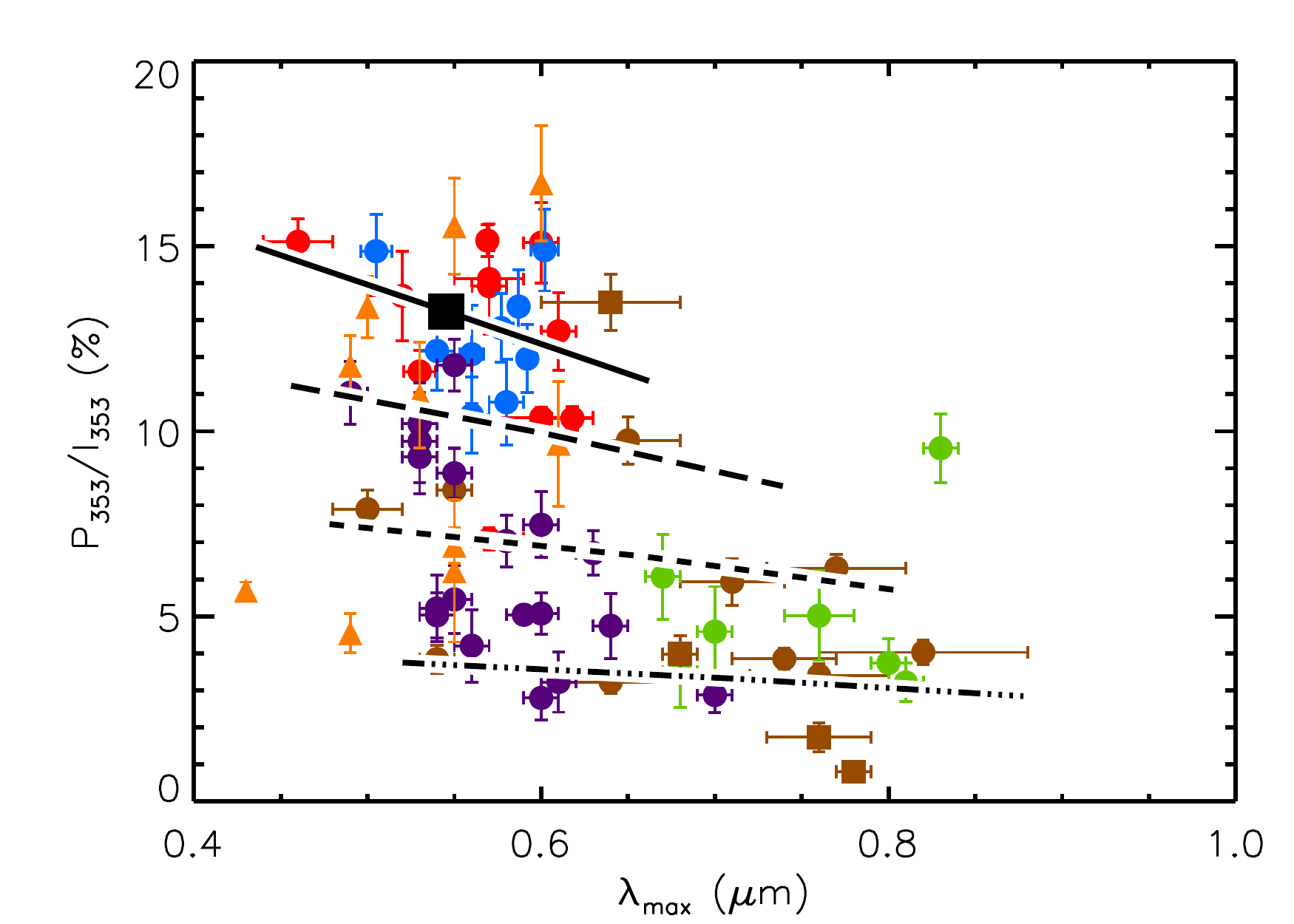}
\includegraphics[width=\hsize]{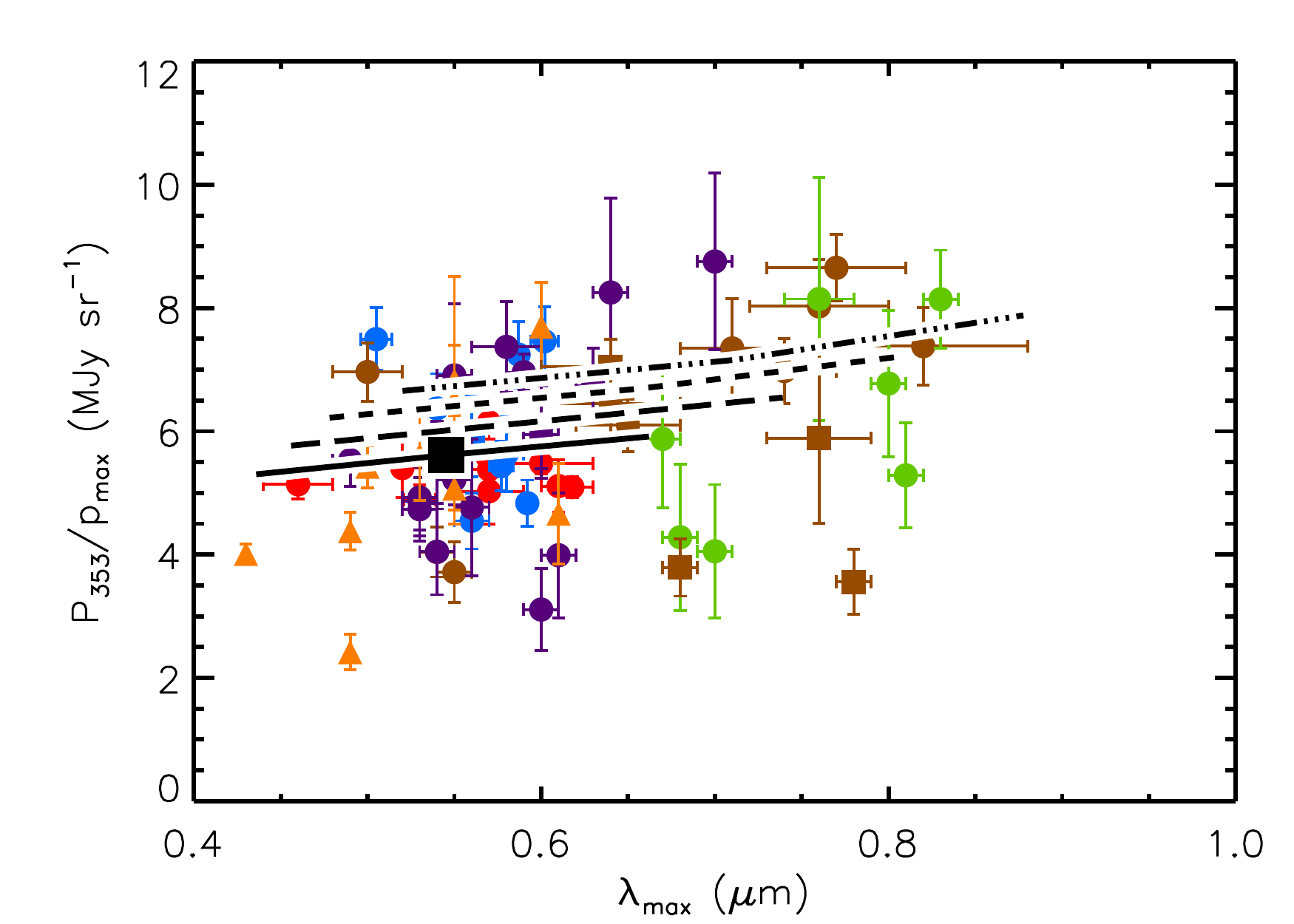}
\caption{
\footnotesize{
Observational data compared with our variable-alignment model ($75 \leq \athresh \leq 150$ nm). The color and symbol scheme are the same as in Fig.~\ref{Obs_Lmax_correlations}. The black square marks the position of the standard model. The top panel does not contain the stars without $\Av$ and $\pv$ measurements.
}
}
\label{Model_Lmax_correlations_athresh}
\end{figure}

The effects of dust alignment efficiency and magnetic field orientation are shown in Fig.~\ref{Model_Lmax_correlations_athresh}, which compares the model results with observational data. Dots represent the observed values of $\pv/\tv$ (top), $\Pem/\Iem$ (middle) and $\Pem/\pmax$ (bottom) as a function of $\lmax$. Curves represent the model; within each curve $\athresh$ varies between 75 and 150 nm and the four curves correspond to the four values of $\gamma$ used, $0^\circ, 30^\circ, 45^\circ$ and $60^\circ$.

The combination of variable alignment and magnetic field orientation can reproduce most of the observations in the case of $\pv/\tv$ and $\Pem/\Iem$, both the general trends and the dispersion.
The curve for $\gamma = 0^\circ$ coincides roughly with the highly polarized, low-$\lmax$ lines of sight in our sample. For higher values of $\gamma$ the polarization decreases, but the dispersion in $\lmax$ caused by the variation of $\athresh$ increases, pushing the maximum $\lmax$ to larger values: as a result, the model predicts that weakly polarized lines of sight that can have either small or large $\lmax$, which is indeed the trend found in the observational data. The relation between $\gamma$ and $\lmax$ described by \cite{Vosh_16} is evident in Fig.~\ref{Model_Lmax_correlations_athresh}: although $\lmax$ is mainly affected by the alignment size threshold $\athresh$, the model curves with a higher $\gamma$ are clearly shifted to higher values of $\lmax$. The figure, however, reveals something more: the strength of the $\gamma$-$\lmax$ relation itself increases with $\athresh$. The leftmost tips of the model curves, corresponding to $\athresh = 75$ nm, all have very similar $\lmax$. On the contrary, the rightmost tips, corresponding to  $\athresh = 150$ nm, show wide differences in $\lmax$, comparable to the differences due to $\athresh$ itself. It should be noted that our model assumes a uniform magnetic field, and therefore it does not include line-of-sight or beam depolarization. If these effects were important, it may mean that the ordered component of the field is closer to the plane of the sky than our model predicts.

Also in Fig.~\ref{Model_Lmax_correlations_athresh} we see that alignment and magnetic field orientation have very little effect on the model results for $\Pem/\pmax$, as was indeed expected: the different curves are close to each other. In fact, the dispersion observed in the observed $\Pem/\pmax$ is much larger than predicted by the model, which suggests that variations in the dust optical properties occur in translucent lines of sight. Again, one possible confounding factor in this interpretation is the depolarization caused by meandering of the magnetic field: we will now attempt to assess its effects.

\begin{figure}
\includegraphics[width=\hsize]{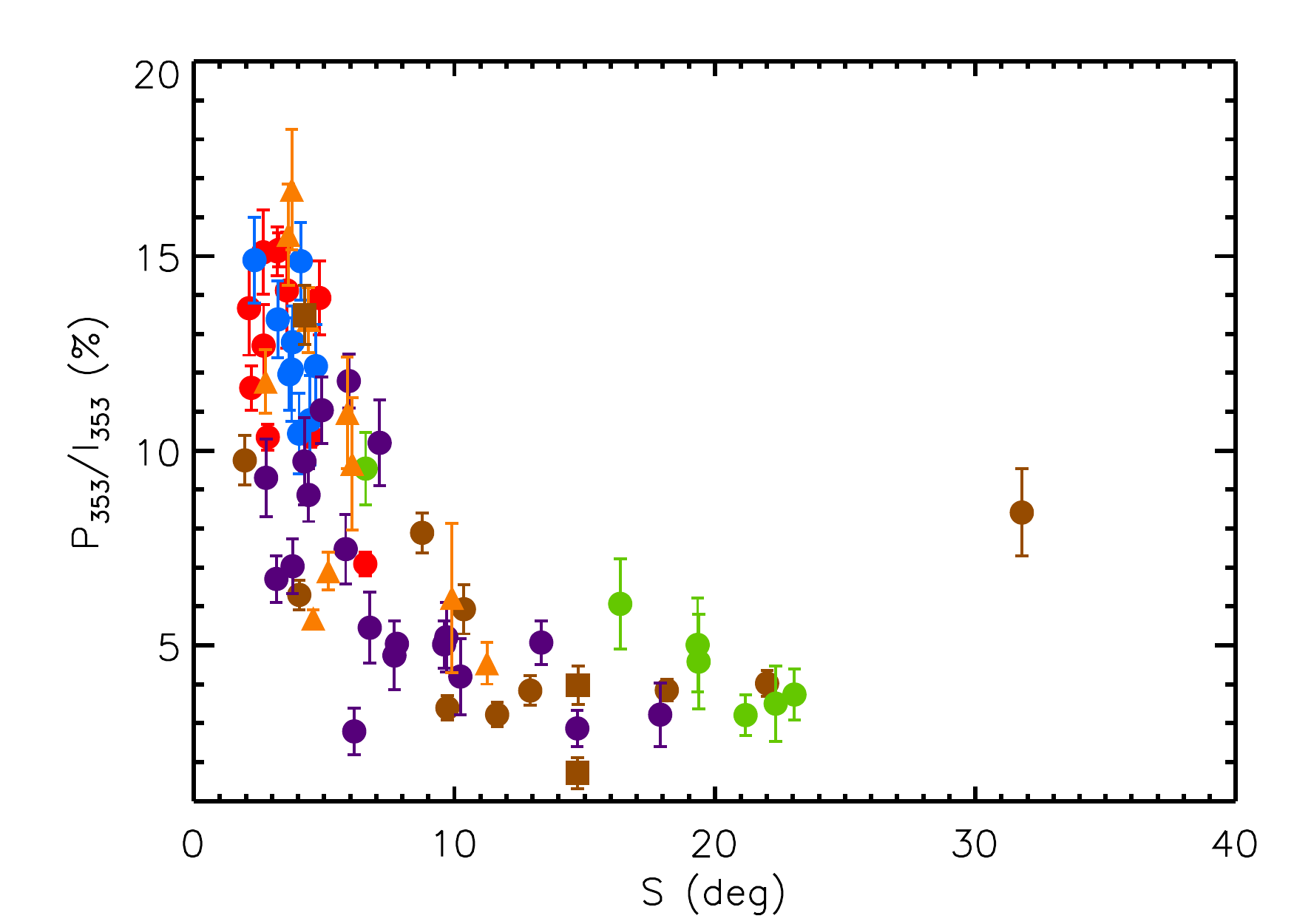}
\includegraphics[width=\hsize]{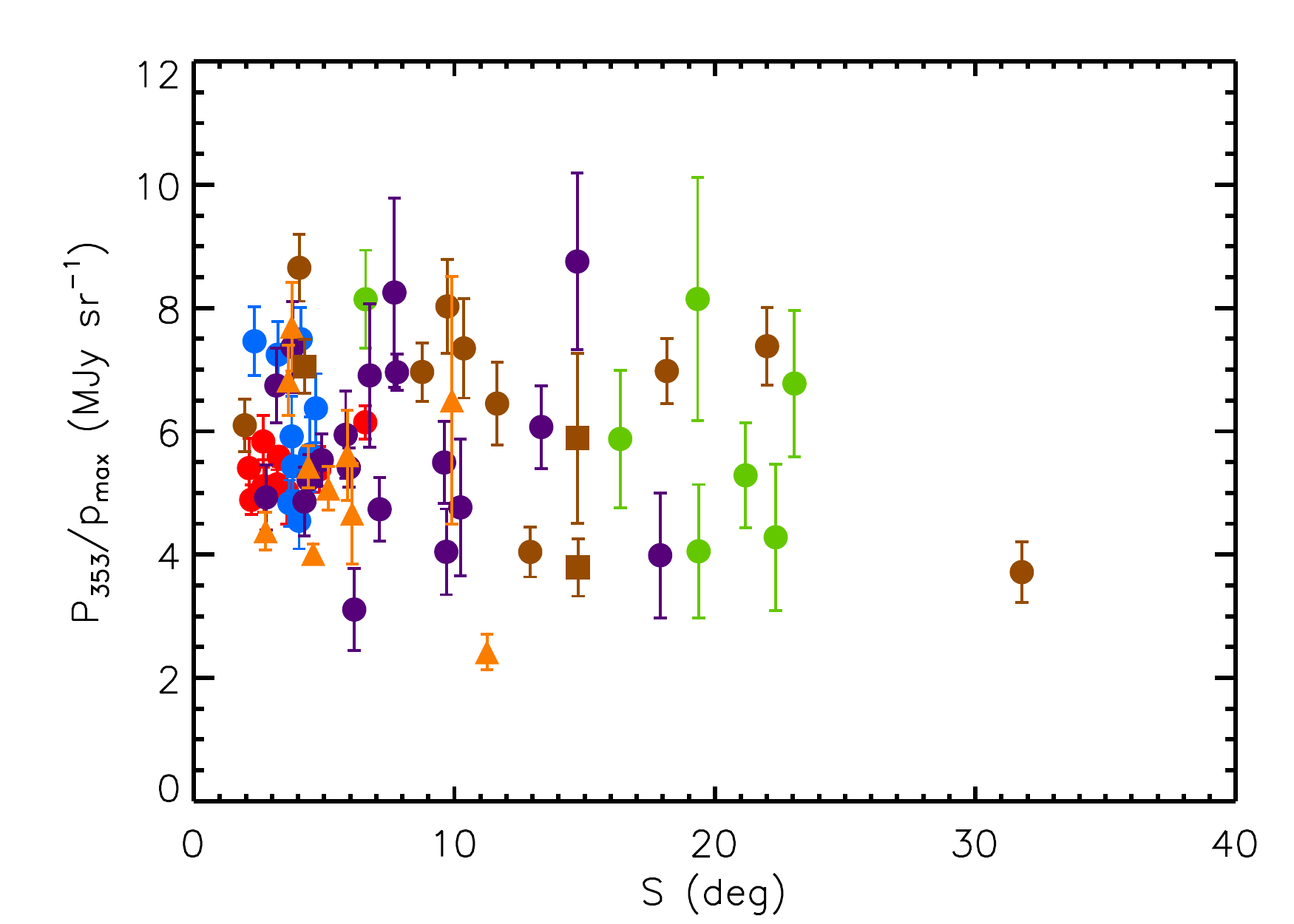}
\caption{
\footnotesize{
The correlation between the observables we studied and the magnetic field meandering as measured by $S$. \textit{Top}: polarization fraction $\Pem/\Iem$ as a function of $\lmax$. \textit{Bottom}: polarization ratio $\Pem/\pmax$ as a function of $\lmax$. The color and symbol scheme are the same as in Fig.~\ref{Obs_Lmax_correlations}.
}
}
\label{Obs_S_Rpp}
\end{figure}
 
While our model assumes a uniform magnetic field, the disorder in the field lines can be estimated from the angle dispersion function $S$ (Section \ref{Section_Data_submm}). In the top panel of Fig.~\ref{Obs_S_Rpp} we see an anticorrelation between $S$ and the polarization fraction in emission, a well-known effect usually attributed to line-of-sight depolarization caused by meandering of the field \citep[\eg][]{Planck_Int_XX}. Beam depolarization is also a possible cause, but we can see that beam effects appear negligible in our sample: the bottom panel of Fig.~\ref{Obs_S_Rpp} shows that there is no clear influence of field meandering on the polarization ratio $\Pem/\pmax$. Beam depolarization, if present, should affect emission but not on extinction, introducing an anti-correlation between $S$ and $\Pem/\pmax$. The fact that we see no such correlation suggests that beam depolarization is negligible in our sample: this supports the idea that we are probing dust in relatively homogeneous regions, which is an important assumption in extinction/emission comparison. The effect of line-of-sight depolarization on the polarization fraction is unfortunately impossible to assess without more advanced modelling, but even if present it should have little effect on $\Pem/\pmax$ following our selection (Sections \ref{Section_Data_Selection} and \ref{Section_Data_depol}). 

\subsection{Grain size distribution}

\begin{figure}
\includegraphics[width=\hsize]{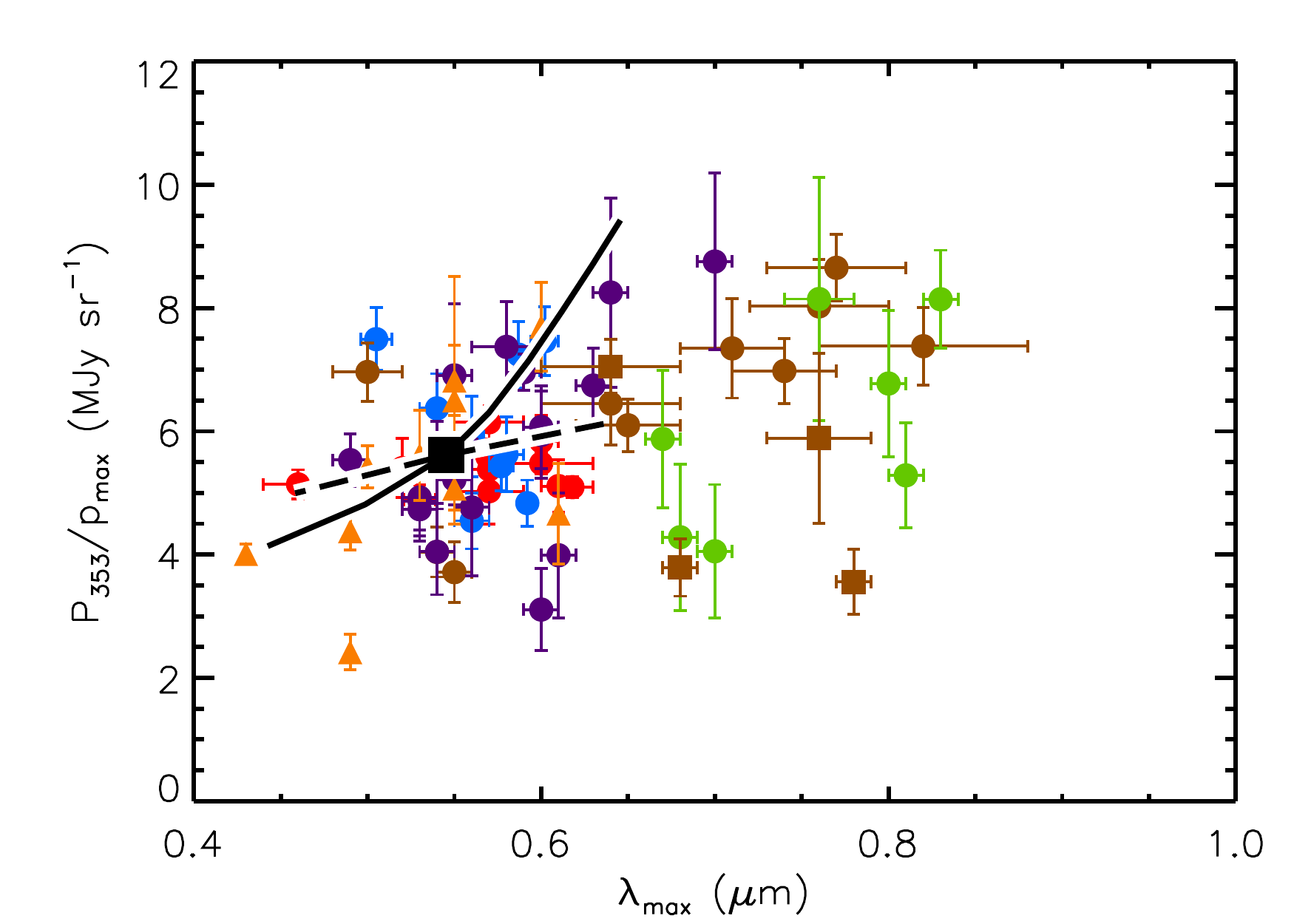}
\caption{
\footnotesize{
The observational polarization ratio $\Pem/\pmax$ compared with a variable size distribution model. The solid line shows the evolution of the model for $350 \leq \amax \leq 10^3$ nm (left to right). The dashed line shows the effect of variable power-law index: $-3.82 \leq \alpha \leq -2.82$ (left to right). The black square marks the position of the standard model. The color and symbol scheme are the same as in Fig.~\ref{Obs_Lmax_correlations}.
}
}
\label{Model_Lmax_correlations_size}
\end{figure}

As explained in Section \ref{Section_model_var_sdist}, we ignore variations of size distribution in carbonaceous grains, which affect $\Iem$ and $\Av$ but not $\Pem$ and $\pmax$. Because of this, our results on the effects of grain growth on observations are unrealistic for $\Pem/\Iem$ and $\pv/\tv$, but realistic for $\Pem/\pmax$, and we show only this last observable. 

Fig.~\ref{Model_Lmax_correlations_size} compares the observations and the model results for $350 \leq \amax \leq 10^3$ nm (solid black line). The effect of a variable $\amax$ on $\lmax$ is small compared to the effect of grain alignment; on the other hand, $\amax$ has a strong effect on $\Pem/\pmax$ and is a plausible contributor to the large dispersion observed in this quantity. For comparison, the picture also shows the model results for fixed $\amax$ and $\alpha$ varying of $\pm 0.5$ around the standard value, \ie\ $-3.82 \leq \alpha \leq -2.82$ (dashed line). The variations in $\alpha$ have a modest effect on $\lmax$ comparable to that of $\amax$; however, unlike $\amax$, the parameter $\alpha$ has nearly no influence on the $\Pem/\pmax$ ratio.

\subsection{Multi-parameter Monte-Carlo}
\label{Section_Results_MC}

\begin{figure}
\includegraphics[width=\hsize]{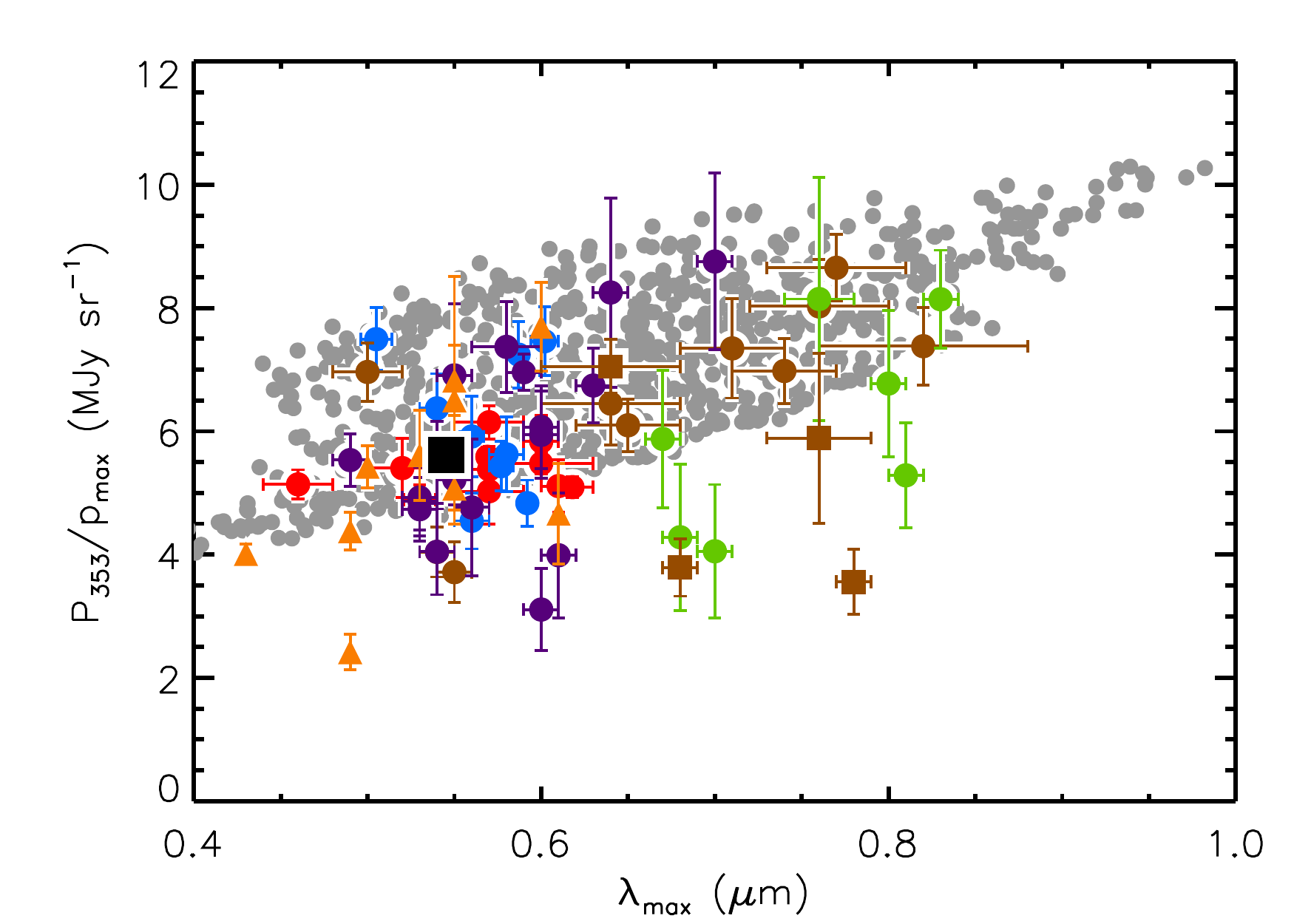}
\caption{
\footnotesize{
The observational polarization ratio $\Pem/\pmax$ compared with Monte-Carlo iterations of the model (grey). The black square marks the position of the standard model. The color and symbol scheme are the same as in Fig.~\ref{Obs_Lmax_correlations}.
}
}
\label{Model_Lmax_correlations_MCsdist}
\end{figure}

We have already mentioned that, realistically, all the model parameters we use will simultaneously vary inside a molecular cloud, and we chose to represent this with a Monte-Carlo simulation (section \ref{Section_Model_MC}). As in the previous section, our modelling does not account for the size distribution of carbon grains and we will only show polarization results. 

Fig.~\ref{Model_Lmax_correlations_MCsdist} compares the Monte-Carlo results to the observational data. The Monte-Carlo was run for  $\gamma = 0^\circ$, $30^\circ$, $45^\circ$ and $60^\circ$; the other model parameters for the Monte-Carlo are uniformly distributed in the regions $75 \, {\rm nm} \leq \athresh \leq 150 \, {\rm nm}$ and $350 \leq \amax \leq 800$ nm. The combined variation of dust alignment, field orientation and grain size allows the model to reproduce most of the observations. However some lines of sight, spanning the full range of observed $\lmax$, have a lower $\Pem/\pmax$ than the model can reproduce. This may be the result of variations in the dust polarization properties due to factors other than size (\eg\ grain shape, porosity, chemical composition), not considered in our model.

\section{Discussion}
\label{Section_Discussion}

\begin{figure}
\includegraphics[width=\hsize]{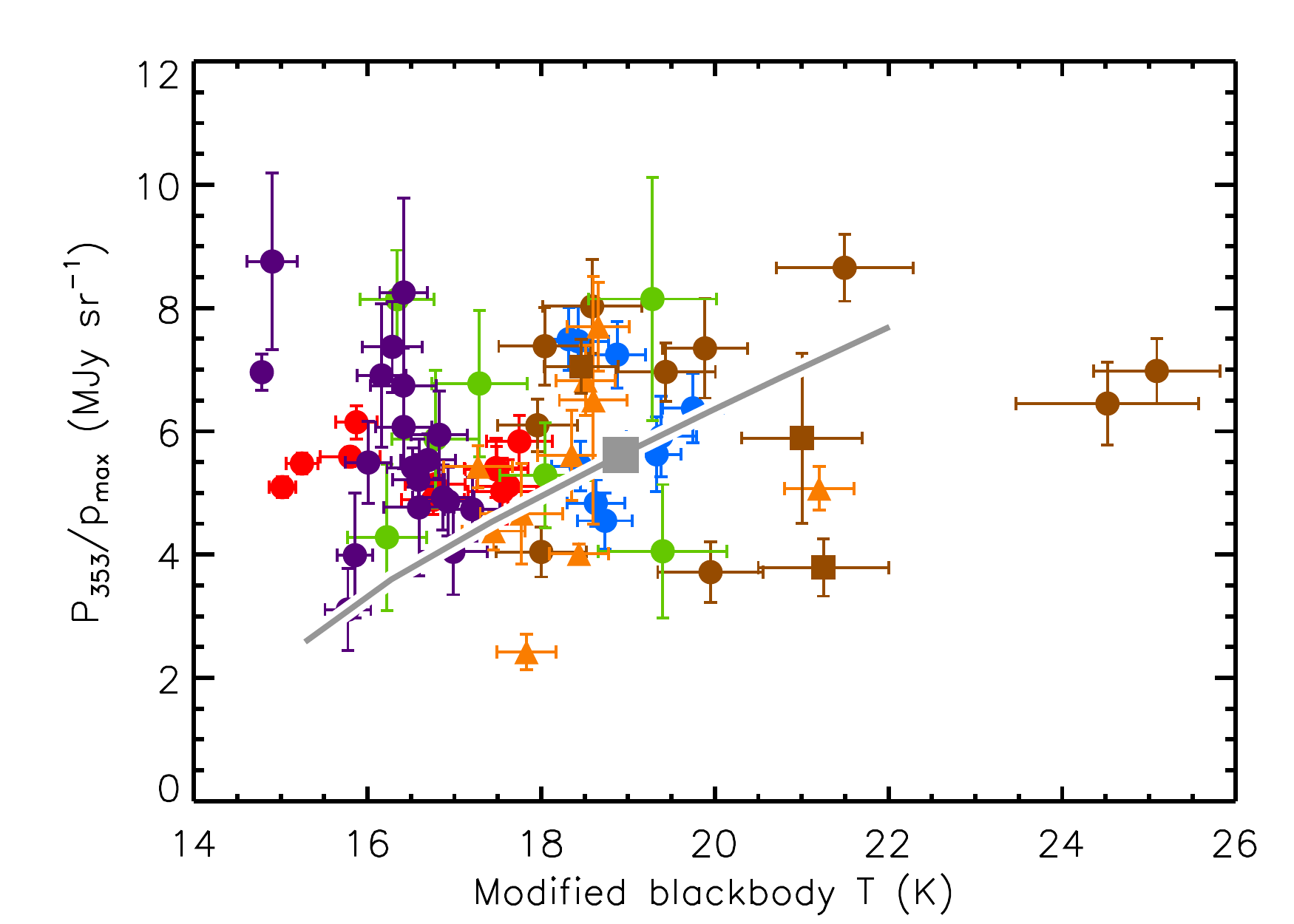}
\includegraphics[width=\hsize]{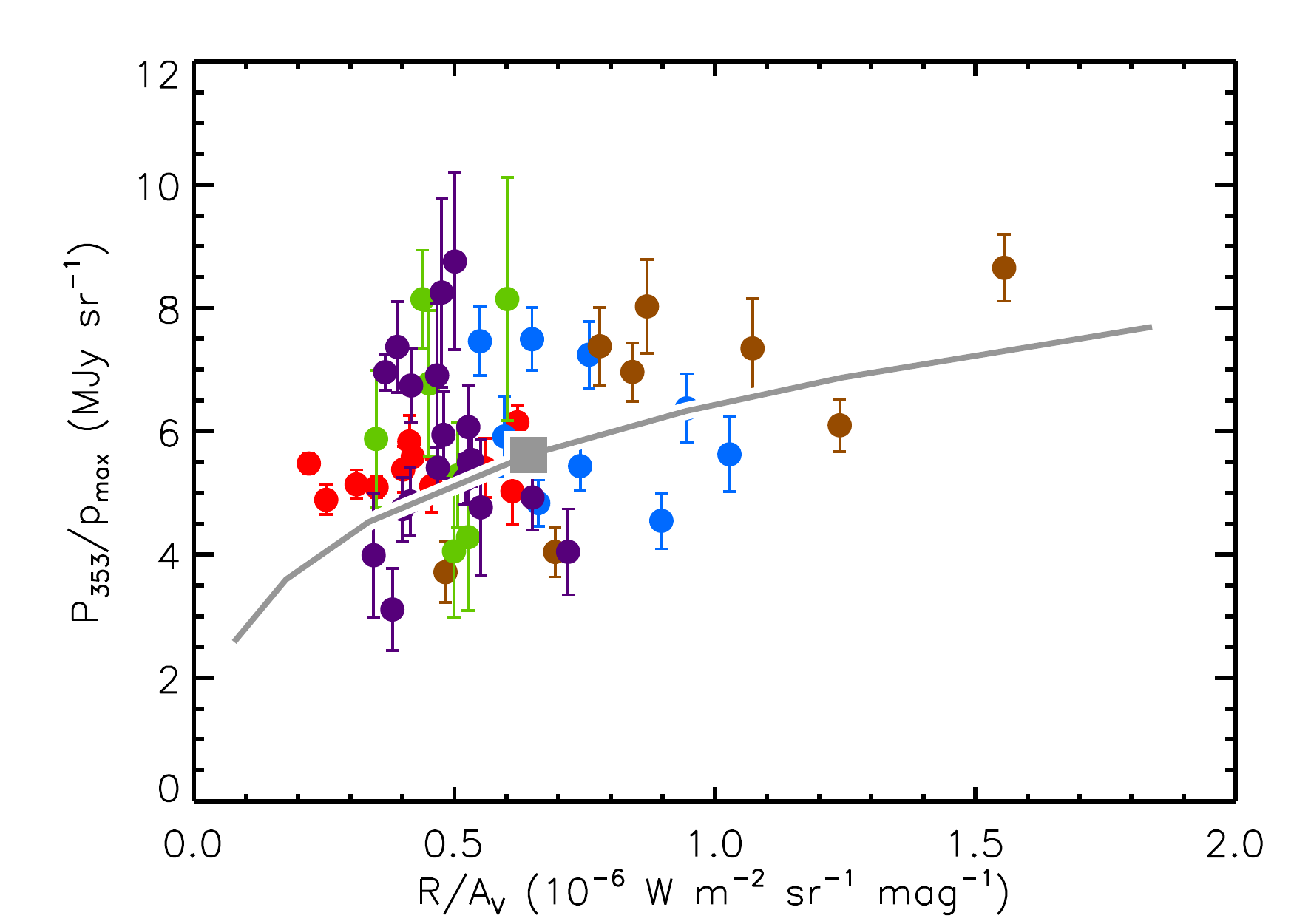}
\caption{
\footnotesize{
The $\Pem/\pmax$ ratio as a function of the modified blackbody temperature (top panel) and as a function of the normalized radiance $\mathcal{R}/\Av$ (bottom panel). The bottom panel does not include stars without $\Av$ and $\pv$ measurements. The grey curves show the model for varying intensity of the radiation field ($0.1 \leq G_0 \leq 3$).The grey square marks the position of the standard model. The color and symbol scheme for the observarions are the same as in Fig.~\ref{Obs_Lmax_correlations}. 
}
}
\label{Model_T_vs_Rpp}
\end{figure}

We have seen that both the grain alignment and the orientation of the magnetic field have important effects on our observables: a combination of variable $\athresh$ and variable $\gamma$ reproduces the general trends and the dispersions for $\pv/\tv$ and $\Pem/\Iem$ as a function of $\lmax$; in fact, we find the familiar ``envelope'' in the distribution of points (Fig.~\ref{Model_Lmax_correlations_athresh}). There are however some stars, mainly in the Musca cloud, with a higher polarization than the model can reproduce. This is because the model was made to reproduce a $\pv/\tv$ of $\sim 3\%$, the usually accepted value for the diffuse interstellar medium \citep[\eg][]{Serkowski, ARAA_review_15} while the outlier stars in Fig.~\ref{Model_Lmax_correlations_athresh} have $\pv/\tv \sim 4\%$. We should point out, however, that the measures in question have relatively large error bars and most stars are within $\sim 1 \sigma$ above the envelope.

We can also see in Fig.~\ref{Model_Lmax_correlations_MCsdist} that a combination of variable alignment, magnetic field orientation and grain growth reproduces the general trend and most of the data scatter in the $\Pem/\pmax$ vs. $\lmax$ relation. Again some data are outside of the model's range; in this case it is the stars with low $\Pem/\pmax$. These lines of sight may belong to regions with different dust properties: while our model reproduces the high $\Pem/\pmax$ ratio from \cite{Planck_Int_XXI}, most models give lower values. Another possibility is that the low $\Pem/\pmax$ is an effect of beam depolarization, which would lower the observed value of $\Pem$ without affecting $\pmax$; however, in Section \ref{Section_Results_Alig+Field} we showed that beam depolarization does not seem important in our sample. A lower $\Pem/\pmax$ is also what we would expect in regions of low dust heating, since $\Pem$ is proportional to dust emission: our model results are calculated for the typical interstellar radiation field intensity\footnote{We used the default interstellar radiation field in DustEM, which is the \cite{Mathis_83} SED for the Solar neighborhood multiplied by the dimensionless intensity parameter $G_0$.} in the diffuse ISM, $G_0 = 1$, but we would expect the radiation field to be less intense in clouds. However, in the long-wavelength range of the Planck function -- which is the case of \Planck\ observations of dust -- emission tends to depend linearly on dust temperature, meaning that the effect of heating on dust SED may be small. 

A proper estimation of heating effects would be no trivial task, as there is no univocal relation between the observed $\Av$ and the extinction actually experienced by dust (see \textbf{AP07} for a discussion), but a preliminary analysis is shown in Fig.~\ref{Model_T_vs_Rpp}. The grey curves in the figure represent the model results for $0.1 \leq G_0 \leq 3$. The top panel shows no visible correlation between $\Pem/\pmax$ and dust temperature, suggesting that $G_0$ only has modest effects \citep[it should be kept in mind, though, that temperature is also influenced by dust evolution: \eg,][]{Stepnik_03, Planck_Int_XVII}. The same conclusion seems supported by the bottom panel: this shows that $\Pem/\pmax$ is not correlated with the normalized radiance $\mathcal{R}/\Av$. Since the radiance $\mathcal{R}$ is the bolometric emission of big dust grains (Section \ref{section_model_fitting}), $\mathcal{R}/\Av$ is a measure of the power emitted (and therefore absorbed) by big grains, averaged on the line of sight.\footnote{Stars 10 and 24 of Ophiuchus \citep[according to the nomenclature of][]{Vrba_93} are not shown in the bottom panel of Fig.~\ref{Model_T_vs_Rpp}, because their large $\mathcal{R}/\Av$ place them outside the plot area. The ($\mathcal{R}/\Av$, $\Pem/\pmax$) values for these outliers are (2.15, 6.45) and (3.24, 6.98) respectively, in the units shown on the figure axes.}

\section{Conclusions and perspectives}
\label{Section_Conclusions}

Dust polarization depends on the efficiency of grain alignment, the orientation and meandering of the magnetic field and the optical properties of the dust itself. Many studies have attempted to explain polarization in molecular clouds in terms of one of these factors (see \eg\ \textbf{AP07} for a study focused on alignment and \cite{Planck_Int_XX} for one focused on magnetic field meandering); nonetheless, it is likely that all factors are at play at once. 

In this paper we use dust model A from \cite{Guillet_17} and vary its alignment efficiency, grain size and magnetic field orientation to reproduce the diverse conditions that one may find in molecular clouds. The results are compared to extinction and emission data from both bibliographic sources and the \Planck\ survey, with particular attention to the polarization observables $\pv/\tv$, $\Pem/\Iem$ and $\Pem/\pmax$ as a function of $\lmax$. 

We find that none of the model parameters employed can explain the full set of observations on its own. Monte-Carlo simulations show that most of the data can be reproduced by letting $\athresh$ vary between 75 and 150 nm, $\amax$ vary between 350 nm and 800 nm, and $\gamma$ vary between $0^\circ$ and $60^\circ$. Thus, any studies of polarization in molecular clouds need to take into account all these aspects to explain the full range of the data. In particular, the ratio $\Pem/\pmax$ is very useful in reducing the contributions of alignment and magnetic field orientation, and highlighting variations in dust properties, and especially size distribution. Within the context of our model, the variations observed in $\Pem/\pmax$ can be partly explained by varying the maximum grain size $\amax$, while the power law index $\alpha$ has little effect.

Nonetheless, some of the observations fall outside the range of the model results, most notably some lines of sight with very low values of $\Pem/\pmax$ that are found over the full range of observed $\lmax$. This is likely to indicate variations in dust properties other than size distribution, such as shape, structure or chemical composition. Non-size-related dust evolution may also influence our estimate of magnetic field orientation: for instance, lines of sight with a low $\Pem/\Iem$ -- which our model attributes to magnetic field lines nearly orthogonal to the plain of the sky (large $\gamma$) -- may be explained instead by dust with a low polarization cross-section. The lines of sight with low $\Pem/\pmax$ can in principle be explained without recurring to dust evolution: a dim radiation field due to extinction, or the beam depolarization due to a disordered magnetic field, would lower the value of polarization in extinction without affecting extinction. However, our analysis conducted in the previous sections suggests that both dust heating and beam depolarization have little effect on our sample.

It is possible that the width of the magnetic field orientation range found by our model ($0^\circ$ to $60^\circ$) is overestimated. Aside from the aforementioned influence of dust properties, field meandering -- which is absent in our model -- could introduce line-of-sight depolarization, the effects of which are degenerate with increasing the angle $\gamma$ of (the ordered component of) the magnetic field. Including field meandering in an ISM model would make for a very interesting follow-up, but a polarized radiation transfer code is needed for that. It would also be useful to independently determine $\gamma$ in future research, \eg\ using MHD simulations or, where available, measures of line-of-sight magnetic field such as Zeeman. 

Not all potentially interesting cases were considered in the present paper: this is a first application of this technique to the study of polarization in molecular clouds. The full implications of this technique will become clearer with more detailed modelling and more observational data. A continuation of this work would benefit from extending the dataset to near-infrared extinction and polarization. Observations in the NIR can probe denser lines of sight than those in the visible; furthermore, increasing the wavelength of observation means getting closer to the Rayleigh limit, so that the observables are better trackers of the overall mass of aligned grains and less dependent on the details of alignment and size distribution. This makes NIR observations an interesting complement to observations in the optical. Finally, different types of observations could improve our constraints on the model variables: maps of molecular lines and elemental depletion could be useful in constraining grain growth processes such as accretion and coagulation.

%

\begin{acknowledgements}
We would like to thank S.~N.~Shore for his stimulating discussion and insightful remarks on a variety of subjects -- from ISM physics to data analysis to writing -- during the editing of this article. We are also indebted to N.~V.~Voshchinnikov for his helpful comments.
\end{acknowledgements}

\bibliographystyle{aa}
\bibliography{biblio2016}

\end{document}